\documentclass[11pt]{article}

\usepackage{amssymb}
\usepackage[dvips]{graphicx}
\usepackage{bm}

\newcommand{\beq}{\begin{equation}}
\newcommand{\eeq}{\end{equation}}
\newcommand{\beqn}{\begin{eqnarray}}
\newcommand{\eeqn}{\end{eqnarray}}

\newcommand{\gsim}{\lower.7ex\hbox{$\;\stackrel{\textstyle>}{\sim}\;$}}
\newcommand{\lsim}{\lower.7ex\hbox{$\;\stackrel{\textstyle<}{\sim}\;$}}

\begin{document}

\title{\bf Derivation of the exact expression for the \boldmath{$D$} function in \boldmath{${\cal N}=1$} SQCD}

\author{
M.  Shifman$^a$ and K. V. Stepanyantz$^b$\\
\\ $^a$ \small{\em William I. Fine Theoretical Physics Institute, University of
Minnesota,}\\
\small{\em Minneapolis, MN 55455, USA}\\
$^b$ \small{\em M.V.Lomonosov Moscow State University, Physical
Faculty,}\\
\small{\em Leninskie Gory, Moscow 119991, Russia}}

\maketitle

\vspace{2mm}

\begin{abstract}

We discuss various details of our derivation of the exact
expression for the Adler $D$ function in ${\cal N}=1$
supersymmetric QCD (SQCD).  This exact formula relates the $D$
function to anomalous dimensions of the matter superfields. Our
perturbative derivation refers to the $D$ function defined in
terms of the bare coupling constant in the case of using the
higher covariant derivative regularization. The exact expression
for this function is obtained by direct summation of supergraphs
to all orders in the non-Abelian coupling constant. As we argued
previously, our formula should be valid beyond perturbation theory
too. The perturbative result we present here coincides with the
general formula order by order. We discuss consequences for ${\cal
N}=1$ SQCD in the conformal window. It is noted that our exact
relation can allow one to determine the (infrared) critical
anomalous dimension of the Seiberg $M$ field present in the dual
theory.

\rule{0mm}{5mm}

 PACS numbers

11.15.-q, 11.30.Pb, 12.60.Jv, 12.38.-t

\end{abstract}

\vfill
\begin{flushright}
FTPI-MINN-15/08\\[1mm]  UMN-TH-3422/15
\end{flushright}

\newpage

\section{Introduction}
\hspace*{\parindent}

In our brief publication \cite{Shifman:2014cya} we reported a new
exact relation between the Adler function $D$ in supersymmetric
QCD (SQCD) with the gauge group $SU(N)$ on the one hand, and the
anomalous dimensions of the matter superfields on the other,
 \beq
 D (Q^2)= \frac{3}{2}  N_c \sum_f
q_f^2\left[  1 -   \gamma \left(\alpha_s (Q^2) \right)\right]\,,
\label{m5}
\eeq
where $N_c$ is the number of colors, $f$ is the flavor index, and $q_f$ is the
corresponding electric charge (in units of $e$). Equation
(\ref{m5}) assumes that all matter fields are in the fundamental
representation of $SU(N)$, although their electric charges can be
different. In calculating  $ \gamma \left(\alpha_s (Q^2) \right)$
one should remember that  $\alpha_s (Q^2) $ runs according to the
Novikov-Vainshtein-Shifman-Zakharov (NSVZ) $\beta$ function
\cite{Novikov:1983uc,Novikov:1985rd,Shifman:1986zi}.

The basic idea of our derivation was the same as that of the NSVZ
$\beta$ functions. The key object of our consideration was the
effective Lagrangian for the external electromagnetic field, with
two terms in it: the $U(1)$ gauge kinetic term and the matter
term. The latter is related to the former through the exact
Konishi anomaly \cite{Konishi:1983hf}, and, therefore, brings in
$\gamma$'s. Although our terminology was perturbative, the result
(\ref{m5}), being based on the exact anomaly relations, should be
exact too. Then we verified that it is valid to all orders in the
$SU(N)$ coupling constant $\alpha_s$ by a direct supergraph
calculation.

From previous works
\cite{Intriligator:2003jj,Kutasov:2003ux,Erkal:2010sh} one could
find that a relation between the two {\em constants }
\begin{equation}
D_*= \frac{3}{2}  N_c \sum_f q_f^2\left[  1 -   \gamma
\left(\alpha_s^* \right)\right]
\end{equation}
is valid in the (super)conformal points, i.e. at the points at
which $\beta_{\rm NSVZ} = 0.$ After our publication
\cite{Shifman:2014cya} it was argued \cite{4} that the above
conformal relation can be generalized for the renormalization
group (RG) flow by considering $R$ charges as functions of the
running $\alpha_s$. The string-based observation \cite{4}
is complementary to our derivation.

Our task in this paper is two-fold. First, we will present a
detailed account of our supergraph calculation, which {\em a
priori} seems quite nontrivial. Second, we will consider
implications of (\ref{m5}) in Seiberg's conformal window
\cite{Seiberg:1994pq,Intriligator:1995au}.

Note, that the Adler function $D$ is directly related
to the celebrated
ratio $R$ defined as
\begin{equation}
R(s) = \frac{\sigma (e^+e^-\to \mbox{ (s)quarks, gluons(inos)} \to  \mbox{hadrons})}{\sigma
(e^+e^-\to \mu^+\mu^-)},
\end{equation}
where $s$ is the (center-of-mass) total energy squared. $R$ plays
a very important role in the QCD-based phenomenology
\cite{Adler:1974gd}. For example, it is used for a precise
determination of the strong coupling $\alpha_s$ from precision
data on $e^+e^-\to$ hadrons in an appropriate range of energy. The
relation between $D$ and  $R$ is as follows \cite{Adler:1974gd}
(see also
\cite{Pivovarov:1991bi,Kataev:1995vh,Radyushkin:1982kg}):
\begin{equation}\label{D_Definition}
D\left(\alpha_s(P^2)\right) \equiv -  12\pi^2\,\frac{d
\Pi(P^2)}{d\log P^2} = P^2 \int\limits_{0}^\infty ds
\frac{R(s)}{(s+P^2)^2},
\end{equation}

\noindent where $\Pi$ denotes the photon polarization operator and
$P$ is the Euclidian momentum. (Throughout this paper Euclidian
momenta are denoted by capital letters.) In our notation
$\Pi$ is related to the inverse invariant charge $d^{-1}$ by the
equation
\begin{equation}
d^{-1} = \alpha_0^{-1} + 4\pi \Pi,
\end{equation}
where $\alpha_0$ is the bare electromagnetic coupling
constant.

To find the Adler function $D$
in SQCD one must
omit all terms proportional to the electromagnetic coupling
constant in calculating the polarization operator $\Pi$. This implies that the electromagnetic field is
considered as an {\em external field}. The $D$ function  encodes
QCD corrections to the photon propagator. In this sense, this
function is similar to ordinary $\beta$ functions. In QCD the Adler
$D$ function was calculated up to the order $O(\alpha_s^4)$
\cite{Baikov:2008jh,Baikov:2010je}. For ${\cal N}=1$ SQCD the
two-loop expression for the Adler function was obtained in
\cite{Kataev:1983at}.

Similarly to the $\beta$ function, the Adler $D$ function  depends
on the subtraction scheme (beyond two loops). In this paper we work
with the $D$  function defined in terms of the bare coupling
constant, namely,
\begin{equation}\label{D_Bare}
D(\alpha_{0s}) \equiv -\frac{3\pi}{2}\, \frac{d}{d\log\Lambda}\,
\alpha_0^{-1}(\alpha,\alpha_s,\Lambda/\mu)\Big|_{\alpha,
\alpha_{s}=\mbox{\scriptsize const}},
\end{equation}
where the derivative is calculated at fixed values of
the renormalized coupling constants
$\alpha(\alpha_0,\alpha_{0s},\Lambda/\mu)$ and
$\alpha_s(\alpha_{0s},\Lambda/\mu)$. Parameter $\Lambda$ denotes the
ultraviolet cut-off which (within the higher
derivative regularization, see below) can be identified with the
dimensionful parameter in the higher derivative regularizing term.

The difference between the definitions of the RG functions in
terms of the bare coupling constant versus renormalized coupling
constant is discussed in detail in \cite{Kataev:2013eta}. In
particular, it was demonstrated that the RG functions defined in
terms of the bare coupling constant depend on the regularization,
but are independent of the subtraction scheme for a fixed
regularization.

\begin{figure}[h]
\vspace*{3.5cm}
\begin{picture}(0,0)
\put(1.7,0.3){\includegraphics[scale=0.8]{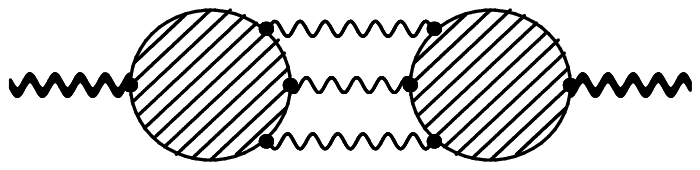}}
\put(2.5,0.1){\mbox{matter loop}} \put(5.5,0.1){\mbox{matter
loop}} \put(7,1.7){\mbox{photon}} \put(1.3,1.7){\mbox{photon}}
\put(2,3.2){\mbox{$SU(N_c)$ gauge}} \put(2,2.7){\mbox{superfiled}}
\put(3.9,2.9){\vector(1,-1){0.9}}
\put(10.3,0.59){\includegraphics[scale=0.41]{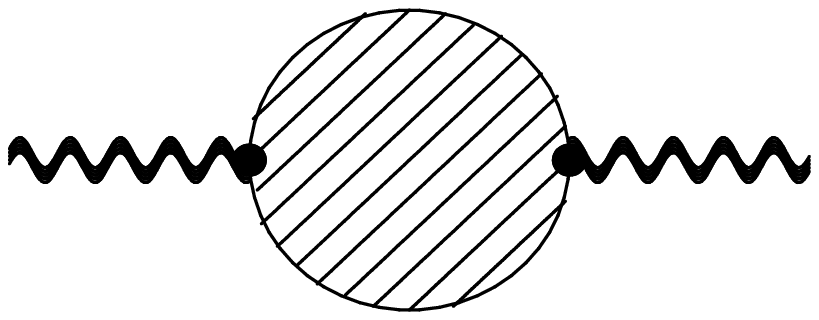}}
\put(11.1,0.1){\mbox{matter loop}} \put(9.6,1.7){\mbox{photon}}
\put(13.2,1.7){\mbox{photon}}
\end{picture}
\caption{Diagrams contributing to the singlet part of the function
$D$ are schematically presented in the left-hand side, and
those contributing to the non-singlet part are presented in the
right-hand side. In the former case the external photon lines are
attached to different loops of the matter superfields, and the
number of internal lines joining the circles should be more than
1. In the latter case the external lines are attached to a single
matter loop.}\label{Figure_Singlet}
\end{figure}

The Adler $D$ function consists of two distinct contributions,

\begin{equation}\label{Singlet}
D(\alpha_s) = \sum\limits_f q_f^2\, D_1(\alpha_s) +
\Big(\sum\limits_f\limits q_f\Big)^2 D_2(\alpha_s),
\end{equation}

\noindent where $q_f$ denotes the ``electric charge" of the flavor
$f$. The first contribution $D_1$ comes from diagrams in which the
external photon lines are attached to the same matter loop, and
the second one $D_2$ (the so-called ``singlet contribution") comes
from diagrams in which the external lines are attached to
different matter loops. The diagrams contributing to both parts of
the function $D$ are sketched in Fig. \ref{Figure_Singlet}.

In this paper we consider ${\cal N}=1$ SQCD with matter
superfields which, in principle, can be in any representation of
the $SU(N)$ gauge group, although Eq. (\ref{m5}) is given for
(anti)fundamental (s)quarks. The matter superfields interact with
the external Abelian gauge superfield (which is a supersymmetric
generalization of the photon field).

Our supergraph derivation of  the exact relation for the $D$
function (\ref{m5}) in terms of $\gamma$'s uses the higher
covariant derivative regularization. In terms of the bare coupling
$\alpha_{0s}$ for each given value of the momentum $Q$ we have
\begin{equation}\label{Exact_D_Function}
D(\alpha_{0s})= \frac{3}{2}  N_c \sum_{f=1}^{N_f} q_f^2\Big( 1 -
\gamma(\alpha_{0s})\Big),
\end{equation}
where $N_c$ is a number of colors, $N_f$ is a number of
flavors, and $\gamma(\alpha_{0s})$ is the anomalous dimension (of
a single chiral matter superfield) defined in terms of the bare
coupling constant

\begin{equation}
\gamma(\alpha_{0s}) \equiv -\frac{d}{d\log\Lambda} \log
Z(\alpha_s,\Lambda/\mu)\Big|_{\alpha_s = \mbox{\scriptsize
const}}.
\end{equation}

\noindent In the two-loop approximation this equation is in
agreement with the result of \cite{Kataev:1983at}

\begin{equation}
D(\alpha_s) =\frac{3}{2} N_c \sum_{f} q_f^2 \left[ 1+
\frac{N_c^2-1}{2N_c} \frac{\alpha_s}{\pi} + O(\alpha_s^2)\right],
\end{equation}

\noindent if we take into account the fact that the one-loop anomalous
dimension is given by the expression

\begin{equation}
\gamma(\alpha_s) = -\frac{N_c^2-1}{2N_c}\frac{\alpha_s}{\pi}
+O(\alpha_s^2).
\end{equation}

\noindent (The two-loop contribution to the Adler function and the
one-loop contribution to the anomalous dimension are
scheme-independent. Consequently, they are the same for the RG
functions defined in terms of the bare coupling constant and the
RG functions defined in terms of the renormalized coupling
constant.)

Certainly, the $D$ function in the supersymmetric case cannot be
applied for the phenomenological purposes, because supersymmetry
is not observed at low energies. Nevertheless, investigation of
SQCD can be useful for better understanding of gauge theories
dynamics. Therefore, our result is promising for
investigating the quantum structure in
supersymmetric gauge theories.

As was mentioned in
\cite{Shifman:2014cya}, the exact relation (\ref{m5})  is closely related to the exact NSVZ
$\beta$ function
\cite{Novikov:1983uc,Novikov:1985rd,Jones:1983ip,Shifman:1986zi}
\begin{equation}\label{NSVZ_Beta}
\beta(\alpha) = - \frac{\alpha^2\Big(3 C_2 - T(R) + C(R)_i{}^j
\gamma_j{}^i(\alpha)/r\Big)}{2\pi(1- C_2\alpha/2\pi)},
\end{equation}
where
\begin{eqnarray}
&& \mbox{tr}\,(T^A T^B) \equiv T(R)\,\delta^{AB};\qquad
(T^A)_i{}^k
(T^A)_k{}^j \equiv C(R)_i{}^j;\qquad\ \nonumber\\[2mm]
&& f^{ACD} f^{BCD} \equiv C_2 \delta^{AB};\qquad\quad r \equiv
\delta_{AA}.\qquad
\end{eqnarray}
The exact NSVZ $\beta$ function relates the renormalization of the
coupling constant in ${\cal N}=1$ supersymmetric theories to the
renormalization of the matter superfields. In the particular case
of $SU(N_c)$ gauge theory with $N_f$ flavors in the fundamental
representation (each flavor gives one Dirac fermion in components)
Eq. (\ref{NSVZ_Beta}) gives
\begin{equation}
\beta(\alpha_s) = - \frac{\alpha^2_s\left(3 N_c - N_f +
N_f\gamma(\alpha_s) \right)}{2\pi(1- N_c \alpha_s/2\pi)}\,,
\label{m6}
\end{equation}
where $\gamma(\alpha_s)$ is the anomalous dimension of the
chiral superfields.

The NSVZ $\beta$ function was originally derived from the
analysis of the structure of instanton corrections, namely,
by requiring their invariance under the renormalization group
\cite{Novikov:1983uc,Novikov:1985rd,Shifman:1999mv}. Another
possibility is to use the structure of the anomaly supermultiplet
\cite{Shifman:1986zi,Jones:1983ip,ArkaniHamed:1997mj}. Yet another (albeit related)
derivation of the NSVZ $\beta$ function was based on the
non-renormalization theorem for the topological term
\cite{Kraus:2002nu}.

Explicit perturbative calculations carried out in dimensional
reduction \cite{Siegel:1979wq} in the $\overline{\mbox{DR}}$
subtraction scheme up to three
\cite{Avdeev:1981ew,Jack:1996vg} and four-loops
\cite{Harlander:2006xq,Jack:2007ni} agree with the NSVZ expression
only in the one- and two-loop approximation. In higher loops the
scheme-dependence of the RG functions becomes essential, and to
obtain the NSVZ $\beta$ function one has to perform a specially tuned
finite renormalization. It was verified in the three- and
four-loop orders that such a finite renormalization exists
\cite{Jack:1996vg,Jack:1996cn,Jack:1997pa,Jack:1998uj}. According
to Ref. \cite{Jack:1996vg} its existence is a non-trivial fact.
The reason is that the NSVZ relation leads to some scheme
independent consequences which should be valid in {\em all}
subtraction schemes \cite{Kataev:2013csa,Kataev:2014gxa}. (The
general equations which describe how the NSVZ $\beta$-function is
changed under finite reparametrizations of the gauge coupling and
finite rescalings of the matter superfields are presented in
\cite{Kutasov:2004xu,Kataev:2014gxa}.)

In the Abelian case the NSVZ scheme (in which the NSVZ relation is
valid in all orders) was constructed in \cite{Kataev:2013eta} by
imposing some simple boundary conditions on the renormalization
constants for ${\cal N}=1$ supersymmetric theories regularized by
higher derivatives. The higher covariant derivative regularization
\cite{Slavnov:1971aw,Slavnov:1972sq} turns out to be very
convenient for investigating quantum corrections in supersymmetric
theories. It is mathematically consistent in contrast with the
dimensional reduction \cite{Siegel:1980qs} and can be formulated
in a manifestly supersymmetric way
\cite{Krivoshchekov:1978xg,West:1985jx}. It can be used in ${\cal
N}=2$ supersymmetric theories too
\cite{Krivoshchekov:1985pq,Buchbinder:2014wra}.

The NSVZ $\beta$ function naturally occurs in ${\cal N}=1$
supersymmetric theories, regularized by higher covariant
derivatives, because momentum integrals for the $\beta$ function
defined in terms of the bare coupling constant are integrals of
total derivatives \cite{Soloshenko:2003nc} and even integrals of
double total derivatives \cite{Smilga:2004zr}. This allows one
to calculate them analytically. Consequently, one obtains
the NSVZ relation for the RG functions defined in terms of the
bare coupling constant.

In the Abelian case this was proved in all
loops \cite{Stepanyantz:2011jy,Stepanyantz:2014ima} and confirmed
by an explicit three-loop calculations
\cite{Stepanyantz:2011jy,Kataev:2013eta,Kataev:2013csa,Kazantsev:2014yna}.
For a generic non-Abelian gauge theory the factorization of relevant
integrals into integrals of double total derivatives was
verified only in the two-loop approximation
\cite{Pimenov:2009hv,Stepanyantz:2011zz,Stepanyantz:2011bz,Stepanyantz:2012zz,Stepanyantz:2012us}.

To prove Eq. (\ref{Exact_D_Function}) we note that the
momentum integrals for the $D$ function (defined in terms of
the bare coupling constant by Eq. (\ref{D_Bare})) in ${\cal N}=1$
SQCD are also integrals of double total derivatives. In this paper
we prove this in all orders using a method similar to the one
proposed in \cite{Stepanyantz:2011jy}. These integrals do not
vanish because of singularities which occur due to the identity
\begin{equation}
\frac{\partial}{\partial Q^\mu} \frac{\partial}{\partial Q_\mu}
\frac{1}{Q^2} = - 4\pi^2 \delta^4(Q).
\end{equation}
(The above equation is written in the Euclidian space.)

Calculating contributions of all singularities we obtain that the
singlet contribution to $D$ symbolically depicted in the left-hand
side of Fig.~\ref{Figure_Singlet} automatically  vanishes once all
relevant supergraphs are summed, while the remaining part of the
$D$ function  satisfies the relation (\ref{Exact_D_Function}).
Thus, highly nontrivial calculation of the Adler function  and the
anomalous dimension and their comparison fully confirms Eqs.
(\ref{m5}) or (\ref{Exact_D_Function}).

The study of the consequences following from Eq. (\ref{m5}) in the conformal window
apparently paves the way to subsequent intriguing explorations.

This paper is organized as follows: In Section \ref{Section_HD} we
explain how one can calculate the Adler $D$ function
using the higher covariant derivative regularization. In
particular, we find a relation between the $D$ function and a part
of the effective action corresponding to the two-point Green
function of the Abelian gauge superfield. The exact formula for
this part of the effective action is constructed in Section
\ref{Section_D_Function}. This formula consists of the three
parts: the singlet contribution (which comes from diagrams in
which external lines are attached to different matter loops), the
contribution of diagrams in which external limes are attached to a
single matter loop, and a non-invariant contribution (due to which
the two-point Green function of the Abelian gauge superfield turns
out to be transversal). Derivation of the exact expression for the
$D$ function by direct summation of supergraphs is presented
in Section \ref{Section_Summation}. First, in Section
\ref{Subsection_Subdiagrams} we find a sum of certain subdiagrams.
Then we substitute this sum to the expression for the $D$ function
using the results of Appendix \ref{Appendix_Subdiagrams}. In
particular, in Section \ref{Subsection_Singlet} we prove that the
singlet contribution to the $D$ function is given by integrals of
double total derivatives and vanishes in all orders. In Section
\ref{Subsection_NonSinglet} we prove that the non-singlet
contribution is also given by integrals of double total
derivatives in the momentum space. However, unlike the integrals
for the singlet contribution, these integrals do not vanish,
because the integrands have singularities. The sum of these
singularities is calculated in Section
\ref{Subsection_Singularities}, where we relate it to the
anomalous dimension of the matter superfields.
Finally, in Section \ref{confwind} we discuss some consequences of the
exact relation (\ref{m5}) for SQCD in the conformal window.

\section{\boldmath{$D$} function and the higher covariant derivative regularization} \hspace*{\parindent}
\label{Section_HD}

Let us consider ${\cal N}=1$ SQCD interacting with the external
Abelian gauge superfield $\mbox{\boldmath$V$}$. This theory can be described by
the action

\begin{eqnarray}
&&\hspace*{-4mm} S = S_{\mbox{\scriptsize gauge}} +
S_{\mbox{\scriptsize matter}} =
\frac{1}{2g_{0}^2}\mbox{tr}\,\mbox{Re} \int d^4x\, d^2\theta\,W^a
W_a + \frac{1}{4 e_0^2} \mbox{Re} \int d^4x\,
d^2\theta\, \mbox{\boldmath$W$}^a \mbox{\boldmath$W$}_a\nonumber\\
&&\hspace*{-4mm} + \sum\limits_{f=1}^{N_f}\Bigg[\frac{1}{4} \int
d^4x\, d^4\theta\Big(\Phi_f^+ e^{2q_f \mbox{\scriptsize
\boldmath$V$}+2V}\Phi_f + \widetilde\Phi_f^+ e^{-2q_f
\mbox{\scriptsize \boldmath$V$}-2V^t} \widetilde\Phi_f\Big) +
\Big(\frac{1}{2}\int d^4x\, d^2\theta\,m_{0f} \widetilde\Phi_f^t
\Phi_f +\mbox{c.c.}\Big)\Bigg],\nonumber\\
\end{eqnarray}

\noindent where $e_0$ is the bare coupling constant for the group
$U(1)$ and $q_f e_0$ is a charge of the superfield with respect to
$U(1)$. The sum runs over all flavors, $a$ is the spinor index,
$g_{0}$ is the bare coupling constant for the non-Abelian gauge
group $G$, which corresponds to the real gauge superfield $V$.
(Also we use the notation $\alpha_0=e_0^2/4\pi$, $\alpha_{0s} =
g_0^2/4\pi$.) The corresponding field strength is
\begin{equation}
W_a \equiv \frac{1}{8} \bar D^2 (e^{-2V} D_a e^{2V}).
\end{equation}
The Abelian gauge field strength $\mbox{\boldmath$W$}_a$
is defined as
\begin{equation}
\mbox{\boldmath$W$}_a = \frac{1}{4} \bar D^2 D_a
\mbox{\boldmath$V$}.
\end{equation}
$m_{0f}$ denotes the bare mass of the matter superfields. Below we
consider only the limit $m_0\to 0$. For simplicity, we will omit
the flavor index $f$. The Abelian gauge superfield
$\mbox{\boldmath$V$}$ is treated as an external field and is
present only in the external lines. Due to loop corrections both
coupling constants of the theory are running. In this paper we
investigate the renormalization of the coupling constant
corresponding to the group $U(1)$ and the exact expression for the
corresponding RG function, which is the Adler $D$ function.

Certainly, for calculating quantum corrections one should
regularize the theory. We are certain that in the Abelian case the NSVZ
$\beta$ function is obtained exactly in all loops for the RG
functions defined in terms of the bare coupling constant if the
theory is regularized by higher derivatives.
For non-Abelian theories the corresponding analysis is not yet fully completed,
but there are multiple indications that that is the case too.
That is why  we
use the higher covariant derivative regularization
\cite{Slavnov:1971aw,Slavnov:1972sq} in this paper, as well as the definition of the RG
functions in terms of the bare coupling constant.

The main idea of the higher derivative regularization is adding to
the classical action a term with the higher derivatives, which
increases a degree of momentum in the propagator. In
supersymmetric theories such terms can be easily constructed by
using ${\cal N}=1$ superfields
\cite{Krivoshchekov:1978xg,West:1985jx,Buchbinder:2014wra}. The
argumentation of this paper does not depend on a particular form
of the higher derivative regularization.\footnote{Even in the
Abelian case there are different versions of the higher derivative
regularization, see, e.g.,
\cite{Pimenov:2009hv,Stepanyantz:2011zz,Stepanyantz:2011bz}, which
lead to the same structure of quantum corrections.}

For definiteness, we will stick to one possible variant of the
higher derivative term. In order to construct it we use the
superfield $\Omega$ which is related to the gauge superfield $V$
as follows:
\begin{equation}
e^{2V} \equiv e^{\Omega^+} e^\Omega\ .
\end{equation}
Under the gauge transformations this superfield
transforms as

\begin{equation}
e^\Omega \to e^{iK} e^\Omega e^{i\lambda}\, ,
\end{equation}

\noindent where $\lambda$ is a chiral superfield,  a
parameter of ordinary gauge transformations, and $K$ is a real
superfield, which reflects an arbitrariness of constructing
$\Omega$ from $V$. Using the superfield $\Omega$ one can construct
the gauge covariant supersymmetric derivatives
\begin{equation}
\nabla_a = e^{-\Omega^+} D_a e^{\Omega^+}; \qquad \bar\nabla_{\dot
a} = e^{\Omega} \bar D_{\dot a} e^{-\Omega}\, .
\end{equation}
(Acting on a superfield $S$ which transforms as $S\to
e^{iK}S$ these derivatives transforms in the same way, $\nabla_a S
\to e^{iK} \nabla_a S$.) Then the possible  higher derivative term
is
\begin{equation}
S_\Lambda = \frac{1}{2g_{0}^2} \mbox{tr}\,\mbox{Re} \int
d^4x\,d^2\theta\, (e^\Omega W^a e^{-\Omega}) \Big[
R\Big(-\frac{\bar\nabla^2 \nabla^2}{16\Lambda^2}\Big)-1\Big]
(e^\Omega W_a e^{-\Omega}),
\end{equation}
where the parameter $\Lambda$ (with dimension of mass)
plays a role of the ultraviolet cutoff. The function $R$ is the
ultraviolet regulator such that $R(0)-1=0$ and $R(x)\to \infty$
for $x\to\infty$. For example, it is convenient to choose
\begin{equation}
R(x) = 1 + x^n\,.
\end{equation}

\noindent In order to fix a gauge it is necessary to add the term
$S_{\mbox{\scriptsize gf}}$ to the action. Also one should
introduce the corresponding ghosts with the action
$S_{\mbox{\scriptsize ghosts}}$. Here we will not concretize these
expressions. We only assume that they do not include matter
superfields $\Phi$ and $\widetilde\Phi$.

It is well known that by introducing the higher derivative term we
regularize all divergences beyond the one-loop approximation
\cite{Faddeev:1980be}. The remaining one-loop divergences (and the
one-loop subdivergencies) should be regularized by inserting the
Pauli-Villars determinants\,\footnote{The Pauli--Villars determinants
should be also introduced for ghosts, but in this paper they are
not essential and we do not write them explicitly.} into the generating functional
\cite{Slavnov:1977zf},
\begin{equation}
\Gamma[\mbox{\boldmath$V$}] = -i \log \int DV D\Phi
D\widetilde\Phi \prod_{I=1}^m
\det(V,\mbox{\boldmath$V$},M_I)^{c_I}
\exp\Big(i(S+S_\Lambda+S_{\mbox{\scriptsize gf}}+
S_{\mbox{\scriptsize ghosts}})\Big),
\end{equation}
where $M_I = a_I \Lambda$ and $a_I$ do not depend on $\alpha_{0s}$
and $\alpha_0$. We are interested in the case $m_{0f}=0$, in
which, for simplicity, it is possible to assume that the
parameters $M_I$ do not depend on the flavor $f$. Note that
sources are not included into this expression, because we consider
only diagrams with the external lines corresponding to the Abelian
superfield $\mbox{\boldmath$V$}$. In order to cancel the remaining
one-loop divergences the coefficients $c_I$ should satisfy the
constraints
\begin{equation}\label{Original_C_I}
\sum\limits_{I=1}^m c_I = 1;\qquad \sum\limits_{I=1}^m c_I M_I^2
=0\,.
\end{equation}

The Pauli--Villars determinants can be presented in the form of
functional integrals over the corresponding Pauli--Villars
superfields
\begin{equation}
\det(V,\mbox{\boldmath$V$},M_I)^{-1} = \int D\Phi_I
D\widetilde\Phi_I e^{iS_I},
\end{equation}
where
\begin{equation}
S_I = \sum\limits_{f}\Bigg[\frac{1}{4} \int d^4x\,
d^4\theta\Big(\Phi_I^+ e^{2q \mbox{\scriptsize
\boldmath$V$}+2V}\Phi_I + \widetilde\Phi_I^+ e^{-2q
\mbox{\scriptsize \boldmath$V$}-2V^t} \widetilde\Phi_I\Big) +
\Big(\frac{1}{2}\int d^4x\, d^2\theta\,M_I \widetilde\Phi_I^t
\Phi_I +\mbox{c.c.}\Big)\Bigg].
\end{equation}
Let us note that the functional integral over the usual
matter fields $\Phi$ and $\widetilde\Phi$ can be also written as a
determinant with $M_0=0$ and $c_0=-1$. This allows to treat the
usual fields and the Pauli--Villars fields in a similar manner and
rewrite Eq. (\ref{Original_C_I}) in a simpler form
\begin{equation}
\sum\limits_{I=0}^m c_I = 0;\qquad \sum\limits_{I=0}^m c_I M_I^2
=0\,.
\end{equation}
The two-point Green function of the Abelian gauge superfield
${\mbox{\boldmath$V$}}$ is transversal
\begin{equation}\label{Two_Point_Function}
\Delta\Gamma^{(2)} = - \frac{1}{16\pi} \int
\frac{d^4p}{(2\pi)^4}\,d^4\theta\,\mbox{\boldmath$V$}(\theta,-p)\,\partial^2\Pi_{1/2}
\mbox{\boldmath$V$}(\theta,p)\,
\Big(d^{-1}(\alpha_0,\alpha_{0s},\Lambda/p)-\alpha_0^{-1}\Big)\,,
\end{equation}
because of the $U(1)$ background gauge invariance. (In
this equation $$\partial^2\Pi_{1/2} = -D^a \bar D^2 D_a/8$$ denotes
the supersymmetric transversal projection operator.) We will
calculate the function
\begin{eqnarray}\label{Adler_Function}
 D(\alpha_{0s}) &=& 6\pi^2 \frac{d}{d\log\Lambda}
\Pi\Big(\alpha_{0s}(\alpha_s,\Lambda/\mu),\Lambda/p\Big)\Big|_{p=0}
\nonumber\\[4mm]
& = & \frac{3\pi}{2} \frac{d}{d\log\Lambda}
\Big(d^{-1}(\alpha_0,\alpha_{0s},\Lambda/p) -
\alpha_0^{-1}\Big)\Big|_{p=0} = \frac{3\pi}{2\alpha_0^2}
\frac{d\alpha_0}{d\log\Lambda}\,,
\end{eqnarray}

\noindent where the differentiation is performed at fixed values
of the renormalized coupling constants $\alpha_s$ and $\alpha$.
Writing the last identity we take into account that the function
$d^{-1}$ expressed in terms of the renormalized coupling constant
should be finite. Possible finite terms proportional to
$p/\Lambda$ vanish in the limit $p\to 0$. In order to extract the
expression (\ref{Adler_Function}) from the effective action we
differentiate $\Delta\Gamma = \Gamma - S$ with respect to
$\log\Lambda$ and then make a substitution
\begin{equation}
\mbox{\boldmath$V$}(x,\theta) \to \theta^4\,.
\end{equation}
Then we easily obtain

\begin{equation}\label{Adler_Function_Final}
\frac{1}{3\pi^2}{\cal V}_4 \cdot D(\alpha_{0s}) =
\frac{d(\Delta\Gamma^{(2)})}{d\log\Lambda}\Big|_{\mbox{\scriptsize
\boldmath$V$}=\theta^4},
\end{equation}

\noindent where ${\cal V}_4\to \infty$ is the (properly
regularized) space-time volume \cite{Stepanyantz:2014ima}.

\section{Exact equation for the two-point function of the Abelian
gauge superfield} \hspace{\parindent}\label{Section_D_Function}

Let us derive the exact expression for the function $D$. First, we
prove that this function is given by integrals of double total
derivatives in the momentum space. This can be done using the
argumentation similar to the one proposed in Ref.
\cite{Stepanyantz:2011jy}. First, it is necessary to calculate
formally the integral over the matter superfields. This can be
done, because the action is quadratic in them. However, this
calculation should be carried out very carefully, because the matter
superfields satisfy the chirality constraint. The result can be
written in the following form \cite{Stepanyantz:2011jy}:

\begin{equation}\label{New_Z}
\exp\Big(i\Gamma[\mbox{\boldmath$V$}]\Big) = \int
DV\,\prod\limits_{I=0}^m \prod\limits_{f=1}^{N_f} \det(\large
\mbox{$\bm{\star}_{I,f}$})^{c_I/2} \exp\Big\{ i
\Big(S_{\mbox{\scriptsize gauge}} + S_\Lambda +
S_{\mbox{\scriptsize gf}} + S_{\mbox{\scriptsize ghosts}}\Big)
\Big\},
\end{equation}

\noindent where we use the notation

\begin{equation}
{\large\mbox{$\bm{\star}$}} \equiv \frac{1}{1-\bm{I_{0}}}; \qquad
\bm{I_0} = \bm{B} P.
\end{equation}

\noindent (For simplicity we omit the subscripts $I$ and $f$ for
${\large \mbox{$\bm{\star}$}}$, $\bm{I_0}$, and $P$, which mark
the dependence on $I$ via the mass $M_I$ and on $f$ via the
electric charge $q_f$.) The matrix

\begin{equation}
\bm{B} \equiv \left(
\begin{array}{cccc}
0 & (e^{2q \mbox{\scriptsize \boldmath$V$}+2V^t}-1) & 0 & 0\\[2mm]
(e^{2q \mbox{\scriptsize \boldmath$V$}+2V}-1) & 0 & 0 & 0\\[2mm]
0 & 0 & 0 & (e^{-2q \mbox{\scriptsize \boldmath$V$}-2V}-1)\\[2mm]
0 & 0 & (e^{-2q \mbox{\scriptsize \boldmath$V$}-2V^t}-1) & 0
\end{array}
\right)
\end{equation}

\noindent encodes vertices of the theory, and the matrix

\begin{equation}
P \equiv \left(
\begin{array}{cccc}
0 & {\displaystyle \frac{\bar D^2 D^2}{16(\partial^2 + M^2)}}
& {\displaystyle \frac{M \bar D^2}{4(\partial^2 + M^2)}} & 0\vphantom{\Bigg(}\\[3mm]
{\displaystyle \frac{D^2 \bar D^2}{16(\partial^2 + M^2)}} & 0 & 0
& {\displaystyle \frac{M D^2}{4(\partial^2 + M^2)}}\vphantom{\Bigg(}\\[3mm]
{\displaystyle \frac{M \bar D^2}{4(\partial^2 + M^2)}}
& 0 & 0 & {\displaystyle \frac{\bar D^2 D^2}{16(\partial^2 + M^2)}} \vphantom{\Bigg(}\\[3mm]
0 & {\displaystyle \frac{M D^2}{4(\partial^2 + M^2)}} &
{\displaystyle \frac{D^2 \bar D^2}{16(\partial^2 + M^2)}} &
0\vphantom{\Bigg(}
\end{array}
\right)
\end{equation}

\noindent contains propagators of various matter superfields. In
our notation the strings and rows of these matrices correspond to
the following sequence of the matter superfields:

\begin{equation}
\Big(\Phi,\ \Phi^*, \widetilde\Phi,\ \widetilde\Phi^*\Big).
\end{equation}

\noindent The expression ${\large \mbox{$\bm{\star}$}}$ encodes a
sequence of vertices and matter propagators. In the case
$\mbox{\boldmath$V$}=0$ we use the notations

\begin{equation}
\star \equiv {\large \mbox{$\bm{\star}$}}\Big|_{\bm{V}=0};\qquad
I_0 \equiv \bm{I_0}\Big|_{\bm{V}=0};\qquad B \equiv
\bm{B}\Big|_{\bm{V}=0}.
\end{equation}

\begin{figure}[h]
\vspace*{9.5cm}

\begin{picture}(0,0)
\put(1,8.5){\Large $\star\quad  =\quad 1\quad +\quad BP\quad
+\quad BPBP\quad +\quad BPBPBP\quad +\quad \ldots$}

\put(5,8){\vector(-1,-2){0.5}} \put(8.1,8){\vector(0,-1){1}}
\put(11.7,8){\vector(0,-1){1}}

\put(3.0,5.15){\includegraphics[scale=0.29]{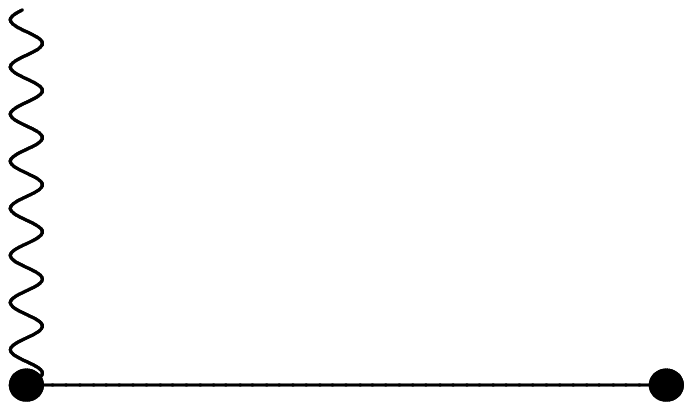}}
\put(6.8,5.1){\includegraphics[scale=0.3]{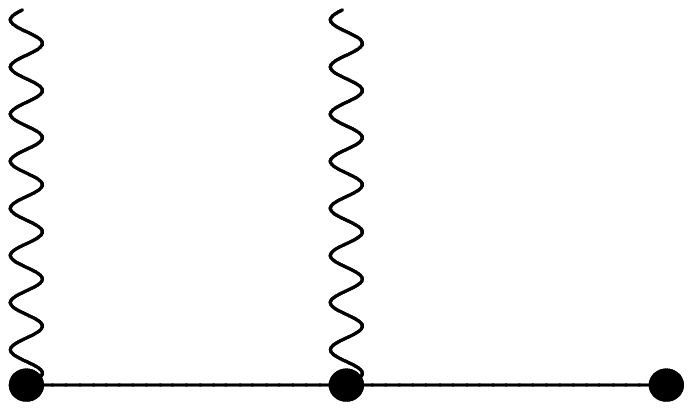}}
\put(10.3,5){\includegraphics[scale=0.44]{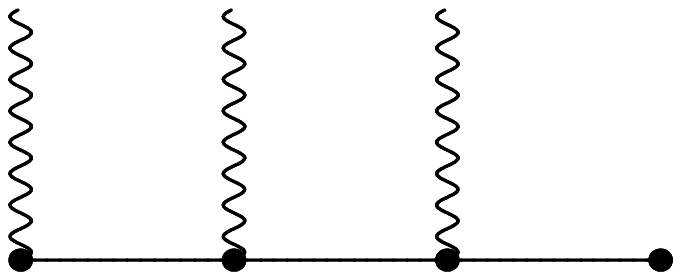}}

\put(2.65,3){\includegraphics[scale=0.33]{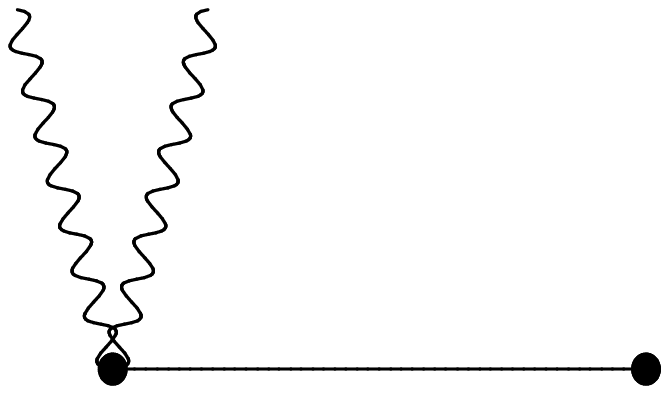}}
\put(6.6,3){\includegraphics[scale=0.4]{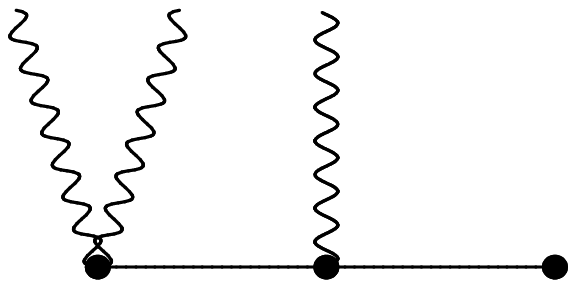}}

\put(2.75,1){\includegraphics[scale=0.33]{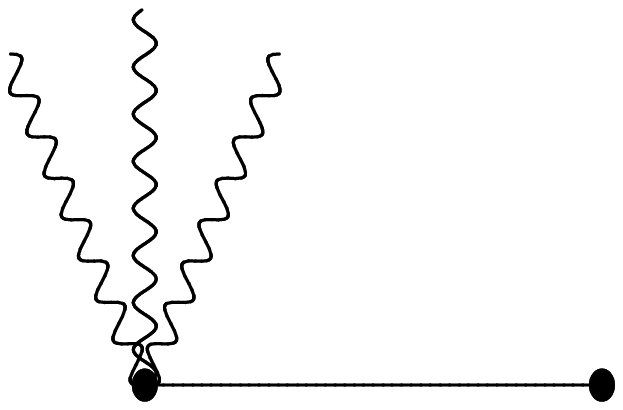}}
\put(6.8,1){\includegraphics[scale=0.4]{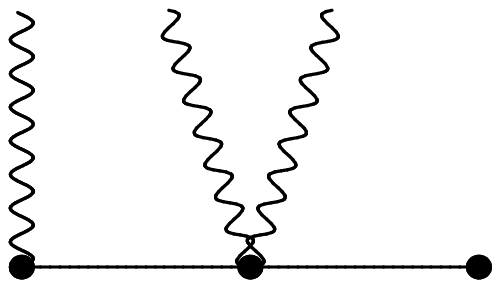}}

\put(11.7,3.6){\Huge $\vdots$}
\put(4.2,-0.3){\Huge $\vdots$}
\put(8.1,-0.3){\Huge $\vdots$}

\end{picture}
\vspace*{0.8cm} \caption{Graphical interpretation of the operator
$\star$.}\label{Figure_Star}
\end{figure}

\noindent These expressions correspond to the diagrams without the
external lines of the Abelian gauge superfield. In particular, the
operator

\begin{equation}
\star = \frac{1}{1-I_0} = 1+ BP + BPBP + BPBPBP+\ldots
\end{equation}

\noindent encodes chains of vertices $B$ (in which the external
gauge superfield $\bm{V}$ is set to zero, and only $V$ is kept) and
the matter propagators $P$. Graphically this equation is presented
in Fig. \ref{Figure_Star}.

Let us recall that the case $I=0$ corresponds to the original
theory in the massless limit, so that $c_0=-1$ and $M_0=0$. Below
we will also use the following operators:

\begin{equation}
(I_1)_a \equiv [I_0,\theta_a];\qquad (\bar I_1)_{\dot a} \equiv
[I_0,\bar\theta_{\dot a}].
\end{equation}

\noindent It is important that all these operators do not
manifestly depend on $\theta$ and $\bar\theta$. (By other words,
$\theta$ and $\bar\theta$ are present only inside the
supersymmetric covariant derivatives.)

In order to calculate the two-point Green function of the
background Abelian gauge superfield $\mbox{\boldmath$V$}$ we
should find terms quadratic in $\mbox{\boldmath$V$}$ in Eq.
(\ref{New_Z}). The result has the form

\begin{eqnarray}\label{Exact_Gamma}
&& \Delta\Gamma^{(2)} = -\frac{i}{2} \Big(\sum\limits_{f=1}^{N_f}
q_f\Big) \cdot \Big\langle \Big(\sum\limits_{I=0}^m c_I
\mbox{Tr}({\bf V} Q J_0 \star)_I\Big)^2
\Big\rangle_{\mbox{\scriptsize 1PI}}
\qquad\nonumber\\
&&\qquad\qquad\qquad +i \sum\limits_{f=1}^{N_f} q_f^2 \cdot
\sum\limits_{I=0}^m c_I \Big\langle \mbox{Tr}({\bf V} Q J_0 \star
{\bf V} Q J_0 \star) + \mbox{Tr}({\bf V}^2 J_0
\star)\Big\rangle_{I,\mbox{\scriptsize 1PI}},\qquad
\end{eqnarray}

\noindent where the symbol 1PI implies that it is necessary to
omit all diagrams except for the one-particle-irreducible (1PI),

\begin{equation}
Q \equiv \left(
\begin{array}{cccc}
1 & 0 & 0 & 0\\[2mm]
0 & 1 & 0 & 0\\[2mm]
0 & 0 & -1 & 0\\[2mm]
0 & 0 & 0 & -1
\end{array}
\right);\qquad J_0 \equiv \left(
\begin{array}{cccc}
0 & e^{2V^t} & 0 & 0\\
e^{2V} & 0 & 0 & 0\\
0 & 0 & 0 & e^{-2V}\\
0 & 0 & e^{-2V^t} & 0
\end{array}
\right) P,
\end{equation}

\noindent and

\begin{eqnarray}\label{Angular_Brackets}
\langle A[V] \rangle \equiv \frac{\displaystyle \int DV\,
A[V]\prod\limits_{I=0}^m \det(\star_{I})^{N_f c_I/2} \exp\Big\{ i
\Big(S_{\mbox{\scriptsize gauge}} + S_\Lambda +
S_{\mbox{\scriptsize gf}} + S_{\mbox{\scriptsize ghosts}} \Big)
\Big\}}{\displaystyle \int DV\,\prod\limits_{I=0}^m
\det(\star_{I})^{N_f c_I/2} \exp\Big\{ i\Big(S_{\mbox{\scriptsize
gauge}} + S_\Lambda + S_{\mbox{\scriptsize gf}} +
S_{\mbox{\scriptsize ghosts}}\Big) \Big\}}.
\end{eqnarray}

\noindent The term in Eq. (\ref{Exact_Gamma}) proportional to
$(\sum_f q_f)^2$ corresponds to attaching the external lines to
different loops of the matter superfields. The terms in Eq.
(\ref{Exact_Gamma}) proportional to $\sum_f q_f^2$ encode diagrams
in which two external lines are attached to a single matter loop.
The exact Adler function $D$ defined in terms of the bare coupling
constant can be found from Eq. (\ref{Exact_Gamma}) by the
following prescription:

\begin{equation}
D(\alpha_{0s}) = \frac{3\pi^2}{{\cal V}_4}\cdot \frac{d\Delta
\Gamma^{(2)}}{d\log\Lambda}\Big|_{\mbox{\scriptsize
\boldmath$V$}=\theta^4}.
\end{equation}

\section{Relation between the function $D$ and the anomalous dimension of the matter superfields.}
\label{Section_Summation}

\subsection{Summation of subdiagrams}
\hspace*{\parindent}\label{Subsection_Subdiagrams}

\begin{figure}[h]
\begin{center}
\includegraphics[scale=0.8]{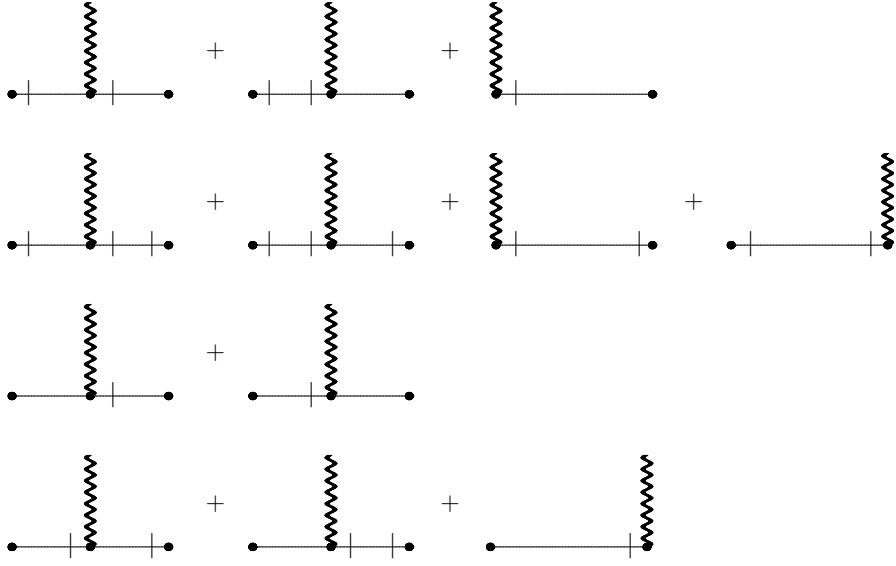}
\end{center}
\vspace*{-1cm} \caption{Summation of subdiagrams. The wavy lines
correspond to the Abelian background gauge superfield
$\mbox{\boldmath$V$}=\theta^4$. To the left and right vertices one
can attach an arbitrary number $\ge 1$ of the internal lines of
the non-Abelian gauge superfield. The middle vertex (if it is
present) does not contain such lines at all.}
\label{Figure_Subdiagrams}
\end{figure}

We calculate the expression (\ref{Exact_Gamma}) after the
substitution $\bm{V}\to \theta^4$. This expression encodes the sum
of all supergraps with two external $\mbox{\boldmath$V$}$ lines.
Summing these supergraphs we encounter certain sequences of
subdiagrams presented in Fig. \ref{Figure_Subdiagrams}, in which
the wavy lines correspond to $\bm{V} = \theta^4$. The left and
right dots correspond to vertices to which an arbitrary number of
$V$ lines can be attached. The middle dot (if it is present)
corresponds to the vertex {\it without} any $V$ lines (and with a
single $\bm{V}$ line). Formally, the subdiagrams presented in Fig.
\ref{Figure_Subdiagrams} can be constructed by transforming the
expression $\star \mbox{\boldmath$V$} J_0$. The details of this
calculation are presented in Appendix \ref{Appendix_Subdiagrams}.

\begin{figure}[h]

\vspace*{3.5cm}
\begin{picture}(0,0)
\put(0.3,2.8){commutation} \put(2.9,2.5){\vector(-1,0){0.9}}
\put(3.3,2.7){$\bm{V}=\theta^4$} \put(7.0,2.7){$\bm{V}=\theta^4$}
\put(12.2,2.7){$\theta$-s} \put(5.4,1.5){$+$} \put(10.3,1.5){$=$}
\put(1.5,0.3){\includegraphics[scale=0.5]{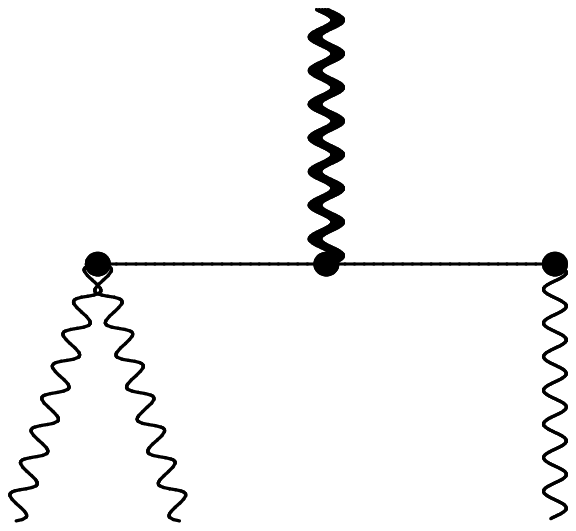}}
\put(6.4,0.2){\includegraphics[scale=0.5]{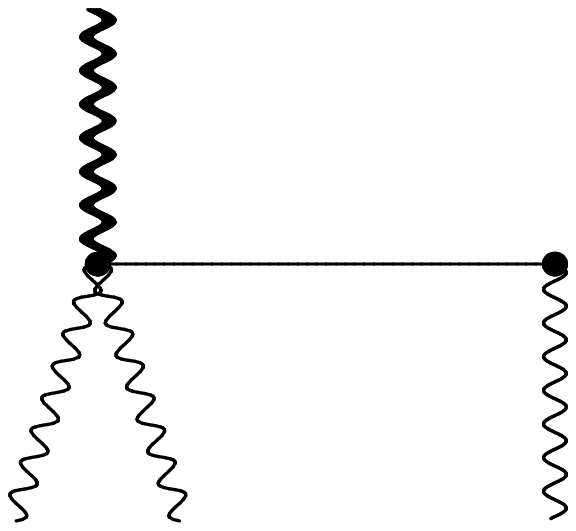}}
\put(11.5,0.2){\includegraphics[scale=0.5]{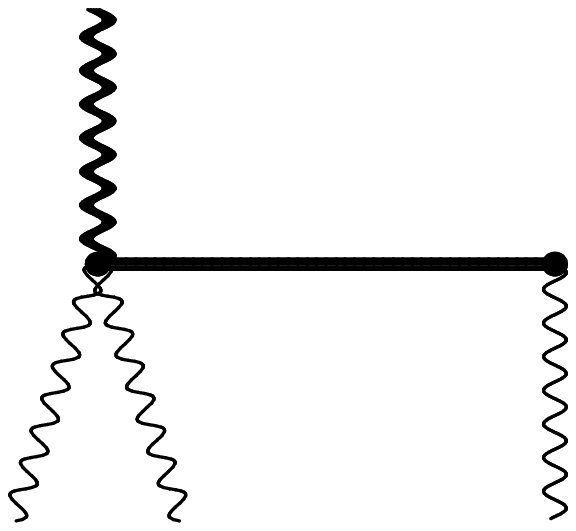}}
\put(13.5,2.9){modified} \put(13.5,2.5){propagator}
\put(13.3,2.4){\vector(-1,-1){0.5}}
\end{picture}

\caption{An example illustrating summation of the effective
diagrams, which gives the modified propagator. (This modified
propagator is denoted by the bold line.)}\label{Figure_Summation}
\end{figure}

To find these sums of subdiagrams the external line is
commuted (with the operators corresponding to the propagators) to
the left. Then we obtain two vertices jointed by the matter line.
This matter line corresponds to a certain operator which is
obtained after the commutation and summing the results. (This
operator replaces the ordinary propagator.) $\theta$'s are
attached to the left vertex. This procedure is qualitatively
illustrated in Fig. \ref{Figure_Summation} for the case corresponding
to the first string in Fig. \ref{Figure_Subdiagrams}.

The results for the sums of the subdiagrams presented in Fig.
\ref{Figure_Subdiagrams} were found in Ref.
\cite{Stepanyantz:2011jy}. (Note that the external line
corresponds to the Abelian field, so that it is possible to use
the result obtained earlier for the Abelian case.) It is
convenient to present the sum of subdiagrams in the matrix form,
namely,

\begin{eqnarray}\label{Subdiagrams} && \hspace*{-5mm}
i\bar\theta^{\dot a} (\gamma^\mu)_{\dot a}{}^b \theta_b [y_\mu^*,
\widetilde Q I_0] - 2\theta^a \theta_a \bar\theta^{\dot b}
[\bar\theta_{\dot b},\widetilde Q I_0] + \widetilde Q\, \bar\theta^{\dot a} B\times\nonumber\\[3mm]
&& \hspace*{-5mm}\left(
\begin{array}{cccc}
0 & 0 & 0 & 0 \\[3mm]
{\displaystyle \frac{i(\gamma^\mu)_{\dot a}{}^b D_b \bar D^2\partial_\mu
}{4(\partial^2 + M^2)^2} + \frac{\bar D_{\dot a}}{\partial^2+M^2}}
& 0 & 0 & {\displaystyle\frac{M \bar D_{\dot a} D^2}{4(\partial^2+M^2)^2}}\\[3mm]
0 & 0 & 0 & 0 \\
0 & {\displaystyle\frac{M \bar D_{\dot a} D^2}{4(\partial^2+M^2)^2}} &
{\displaystyle \frac{i(\gamma^\mu)_{\dot a}{}^b D_b \bar D^2\partial_\mu
}{4(\partial^2 + M^2)^2} + \frac{\bar D_{\dot a}}{\partial^2+M^2}} & 0
\end{array}
\right)\nonumber\\[3mm]
&& \hspace*{-5mm} +\mbox{terms without
$\bar\theta$},\vphantom{\Big(}
\end{eqnarray}

\noindent
where $y_\mu^* = x_\mu - i \bar\theta^{\dot a} (\gamma_\mu)_{\dot a}{}^b \theta_b$ and

\begin{equation}\label{Tilde_Q}
\widetilde Q \equiv \left(
\begin{array}{cccc}
-1 & 0 & 0 & 0\\[2mm]
0 & 1 & 0 & 0\\[2mm]
0 & 0 & 1 & 0\\[2mm]
0 & 0 & 0 & -1
\end{array}
\right).
\end{equation}

\noindent Eq. (\ref{Subdiagrams}) contains exponents corresponding
to the left vertices and the (modified) propagators which are
attached to these vertices from the right. Below we will use the
following properties of the matrix (\ref{Tilde_Q}):

\begin{equation}
[\widetilde Q,I_0]=0;\qquad [\widetilde Q, \star] = 0;\qquad
\widetilde Q^2 = 1.
\end{equation}

\subsection{External lines are attached to different matter loops}
\hspace*{\parindent}\label{Subsection_Two_Matter_Loops}\label{Subsection_Singlet}

Let us consider the term in Eq. (\ref{Exact_Gamma}) proportional
to $(\sum_f q_f)^2$. It corresponds to the case in which the
external lines are attached to different closed loops of the
matter superfields and, therefore, contributes to the singlet part
of the function $D$. Such diagrams are sketched in the left-hand
side of Fig. \ref{Figure_Singlet}. The term under consideration
contain the expression

\begin{equation}\label{Single_Trace}
\mbox{Tr}(\mbox{\boldmath$V$} QJ_0\star)
\end{equation}

\noindent in the second power. In Appendix
\ref{Appendix_Subdiagrams} we prove that this expression can be
presented as

\begin{equation}\label{Single_Trace2}
\mbox{Tr}(\mbox{\boldmath$V$} QJ_0\star) =
\mbox{Tr}\Big\{\star\Big(BP (\bm{V} Q) B_0 P + B (\bm{V} Q) (\Pi_+
P) + B(P\Pi_-) (\bm{V}Q) \Big) \Big\},
\end{equation}

\noindent where $\Pi_\pm$ are chiral projection operators (given
by Eq. (\ref{Chiral_Projectors})) and $B_0$ (given by Eq.
(\ref{B0})) corresponds to the vertex with a single external
$\bm{V}$ line and no $V$ lines. The expression in the round
brackets coincides with the sum of subdiagrams presented in Fig.
\ref{Figure_Subdiagrams}, which was calculated in
Section \ref{Subsection_Subdiagrams} and is given by Eq. (\ref{Subdiagrams}).

It is well known that any supergraph does not vanish only if it
contains $\theta^4$. Therefore, if we multiply two expressions
(\ref{Single_Trace2}), the non-trivial contributions come only
from terms which are linear in $\bar\theta$. (There are no terms
quadratic in $\bar\theta$ in Eq. (\ref{Subdiagrams}).) Let us
consider the sum of the diagrams in which the  matter loop under consideration
(corresponding to the expression (\ref{Single_Trace})) has $n$
vertices with the internal gauge lines. (Vertices with a single
external \mbox{\boldmath$V$} line are not summed.) We will denote
the corresponding contribution as

\begin{equation}\label{Single_Trace_N}
\mbox{Tr}(\mbox{\boldmath$V$} QJ_0\star)_n\,.
\end{equation}

\noindent Using Eqs. (\ref{Single_Trace2}) and (\ref{Subdiagrams})
after the substitution $\mbox{\boldmath$V$}\to \theta^4$ this
expression can be presented as

\begin{eqnarray}
&& \mbox{Tr}(\theta^4 QJ_0\star)_n = \mbox{Tr}\Big( i\bar\theta^{\dot c} (\gamma^\mu)_{\dot c}{}^d \theta_d
[y_\mu^*,\widetilde QI_0] \star -2 \theta^c\theta_c \bar\theta^{\dot d}
[\bar\theta_{\dot d},\widetilde Q I_0] \star +\mbox{$\bar\theta^1$
terms}\Big)_n\qquad \nonumber\\
&& = \mbox{Tr}\Big( i\bar\theta^{\dot c} (\gamma^\mu)_{\dot c}{}^d \theta_d
[y_\mu^*,\widetilde QI_0] (\star)_{n-1} -2 \theta^c\theta_c \bar\theta^{\dot d}
[\bar\theta_{\dot d},\widetilde Q I_0] (\star)_{n-1} +\mbox{$\bar\theta^1$
terms}\Big).
\end{eqnarray}

\noindent (One vertex is written explicitly and corresponds to $B$
inside $I_0$. Therefore, $\star$ should give $n-1$ vertices.)

Comparing the Taylor expansions one can easily see that

\begin{equation}\label{Squared_Star}
(\star^2)_n = (n+1) (\star)_n\,.
\end{equation}

\noindent
Therefore, after a cyclic permulation the  expression under consideration can be written as
follows:
\begin{eqnarray}
&& \frac{1}{n} \mbox{Tr}\Big\{ \widetilde Q\Big(-2\theta^c\theta_c
\bar\theta^{\dot d} \star [\bar\theta_{\dot d},I_0] \star +
i\bar\theta^{\dot c} (\gamma^\mu)_{\dot c}{}^d \theta_d \star
[y_\mu^*,I_0] \star\Big)
+\mbox{$\theta^2$,$\bar\theta^1$,$\theta^1$,$\theta^0$
terms}\Big\}_n \quad\nonumber\\[3mm]
& =& \frac{1}{n}\mbox{Tr}\Big\{ \widetilde Q\Big(-
2\theta^c\theta_c \bar\theta^{\dot d} [\bar\theta_{\dot d},\star] +
i\bar\theta^{\dot c} (\gamma^\mu)_{\dot c}{}^d \theta_d [y_\mu^*,\star]\Big)
+\mbox{$\theta^2$,$\bar\theta^1$,$\theta^1$,$\theta^0$
terms}\Big\}_n\,.
\end{eqnarray}

\noindent
Again comparing the Taylor expansions we obtain

\begin{equation}
(\log \star)_n = \frac{1}{n} (\star)_n\, .
\end{equation}

\noindent Therefore, Eq. (\ref{Single_Trace_N}) gives
\begin{eqnarray}
&& \mbox{Tr}\Big\{ \widetilde Q\Big(-2\theta^c\theta_c
\bar\theta^{\dot d} [\bar\theta_{\dot d}, \log \star ] + i
\bar\theta^{\dot c} (\gamma^\nu)_{\dot c}{}^d \theta_d
[y_\nu^*,\log
\star]\Big)+\mbox{$\theta^2$,$\bar\theta^1$,$\theta^1$,$\theta^0$
terms}\Big\}_n\nonumber\\[3mm]
&& = \mbox{Tr}\Big\{  i \bar\theta^{\dot c} (\gamma^\nu)_{\dot
c}{}^d \theta_d\, \widetilde Q\, [x_\nu,\log
\star]+\mbox{$\theta^2$,$\bar\theta^1$,$\theta^1$,$\theta^0$
terms}\Big\}_n\,,
\end{eqnarray}

\noindent where we take into account the fact that the trace of
$\theta$ commutators always gives 0. Again using this fact and
calculating the square of the last expression we see that the
remaining terms proportional to $\theta^2$, $\bar\theta^1$,
$\theta^1$, or $\theta^0$ do not give $\theta^4$ and can be
omitted. Therefore, (summing contributions for all $n$) we arrive
at

\begin{equation}\label{Single_Trace_Final}
\mbox{Tr}(\mbox{\boldmath$V$} QJ_0\star)\Big|_{\mbox{\scriptsize
\boldmath$V$}=\theta^4} \to\ i \mbox{Tr}\, \bar\theta^{\dot c}
(\gamma^\nu)_{\dot c}{}^d \theta_d\, \widetilde Q\, [x_\nu,\log
\star]\, ,
\end{equation}

\noindent where the symbol $\to$ means that in the right hand side
we omit terms which give vanishing contribution to the whole
supergraph.

The total singlet contribution to the effective action can be
written as

\begin{equation}\label{Singlet_Delta_Gamma}
\Delta\Gamma^{(2)} = \frac{i}{2}\Big(\sum\limits_{f=1}^{N_f}
q_f\Big)^2\cdot \Big\langle \Big(\sum\limits_{I=0}^m c_I\,
\mbox{Tr}\, \bar\theta^{\dot a} (\gamma^\nu)_{\dot a}{}^b
\theta_b\, \widetilde Q\, [x_\nu,\log(\star_I)]\Big)^2\Big\rangle.
\end{equation}

The commutator with $x^\mu$ in the momentum space gives an
integral of a total derivative. Taking into account that these
commutators enter Eq. (\ref{Singlet_Delta_Gamma}) in the second
power, we see that the contributions of the diagrams under
consideration (in which external lines are attached to distinct
loops of the matter superfields) are given by integrals of double
total derivatives.

Each of these
total derivatives is taken with respect to the momentum of its
closed matter loop. It is important that in this case no
singularities appear in the integrand, because there are only factors $Q^{-2}$ in the
numerator, and there are no factors $Q^\mu/Q^4$. Therefore, all integrals
of total derivatives vanish.

Thus, the class of diagrams
considered in this section gives vanishing contribution to the
Adler  $D$ function.

Let us also note that if the matter loop corresponds to the usual
superfields $\Phi$ and $\widetilde\Phi$ (for which $M=0$), Eq.
(\ref{Single_Trace_Final}) can be equivalently rewritten as

\begin{equation}
\mbox{Tr}(\mbox{\boldmath$V$} QJ_0\star)\Big|_{\mbox{\scriptsize
\boldmath$V$}=\theta^4} \to\ 2i\, \mbox{Tr}\, \bar\theta^{\dot c}
(\gamma^\nu)_{\dot c}{}^d \theta_d\, [x_\nu,\log(*) -
\log(\widetilde *)]\,,
\end{equation}

\noindent where

\begin{equation}
* \equiv \frac{1}{1-(e^{2V}-1)\bar D^2 D^2/16\partial^2};\qquad
\widetilde
* = \frac{1}{1-(e^{-2V^t}-1)\bar D^2 D^2/16\partial^2}\,.
\end{equation}

\noindent Therefore, taking into account that $c_0=-1$ the result
can be also presented in the form

\begin{equation}
\Delta\Gamma^{(2)} = 2i\Big(\sum\limits_{f=1}^{N_f}
q_f\Big)^2\cdot \Big\langle \Big[ \mbox{Tr}\, \bar\theta^{\dot c}
(\gamma^\nu)_{\dot c}{}^d \theta_d [x_\nu,\log(*) -
\log(\widetilde *)] + (PV)\Big]^2 \Big\rangle = 0,
\end{equation}

\noindent which was used in \cite{Shifman:2014cya}. (Here $(PV)$
denotes contributions of diagrams with the loops of the
Pauli--Villars superfields.) As was discussed above, the sum of
the diagrams with the Pauli-Villars loop(s) is also given by a
vanishing integral of a double total derivative.

\subsection{External lines are attached to a single matter loop}
\hspace*{\parindent}\label{Subsection_One_Matter_Loop}\label{Subsection_NonSinglet}

In this case we should consider the terms in Eq.
(\ref{Exact_Gamma}) proportional to $\sum_f q_f^2$, which give the
non-singlet contribution. The term containing $\bm{V}^2$ vanishes
after the substitution $\bm{V}\to \theta^4$. The remaining term

\begin{equation}\label{Non-Singlet_Term}
i \sum\limits_{f=1}^{N_f} q_f^2 \cdot \sum\limits_{I=0}^m c_I
\Big\langle \mbox{Tr} \left(\bm{V} Q J_0 \star \bm{V} Q J_0 \star
\right) \Big\rangle_{I,\mbox{\scriptsize 1PI}}
\end{equation}

\noindent gives six effective diagrams with two external lines
corresponding to the Abelian background superfield. These diagrams
are presented in Fig. \ref{Figure_Effective_Diagrams}. They can be
obtained from Eq. (\ref{Non-Singlet_Term}) using the results of
Appendix \ref{Appendix_Subdiagrams}. The first five diagrams $(a)$
-- $(e)$ contain two group of subdiagrams presented earlier in
Fig. \ref{Figure_Subdiagrams}. Each bold line corresponds to the
modified propagator which appears after summation of the
corresponding group of subdiagrams (see Fig.
\ref{Figure_Summation}). Expressions for the external lines are
obtained by extracting the terms proportional to the corresponding
$\theta$ structure from Eq. (\ref{Subdiagrams}). There are no
other relevant diagrams, because a non-vanishing supergraph should
contain the second power of $\theta$ and the second power of
$\bar\theta$. (The diagram $(d)$ contains $\theta^2$, because one
$\theta$ comes from the commutator with $y_\mu^*$.)

The last effective diagram $(f)$ contains contribution of those
graphs in which external lines are close to each other. In
Appendix \ref{Appendix_Subdiagrams} we explain why they should be
added. The vertex in the diagram $(f)$ corresponds to the sum of
many subdiagrams. All of them are collected in Ref.
\cite{Stepanyantz:2011jy}. Their structure is described in
Appendix \ref{Appendix_Subdiagrams}.

\begin{figure}[h]
\hspace*{3mm}
\includegraphics{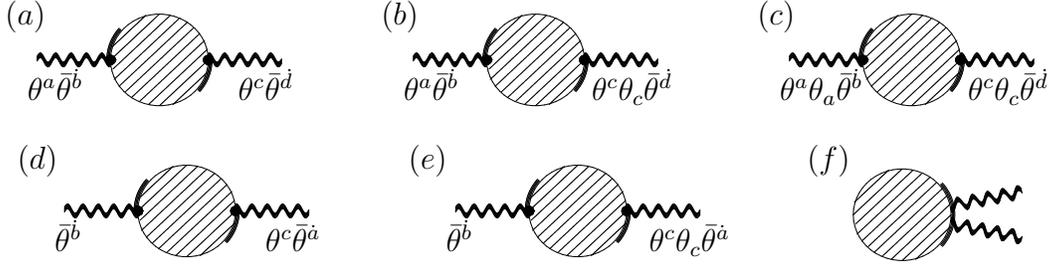}
\caption{Effective diagrams which appear if the external
$\mbox{\boldmath$V$}$ lines are attached to a single matter loop.}
\label{Figure_Effective_Diagrams}
\end{figure}

Calculation of the graphs $(a)$ -- $(f)$ exactly repeats the
corresponding calculation which was carried out in Ref.
\cite{Stepanyantz:2011jy} in the Abelian case. The only difference
is the presence of the charges $q_f^2$ and the sum over all
flavors. All non-Abelian effects are encoded in the angular
brackets, which are now defined in a different way (due to the
different form of the gauge part of the action). The result can be
written in the following form:\footnote{Unlike Ref.
\cite{Stepanyantz:2011jy} here we treat the usual matter
superfields and the Pauli--Villars superfields in the same way.}

\begin{eqnarray}\label{Double_Derivative}
&& i \frac{d}{d\log\Lambda} \sum\limits_f q_f^2
\sum\limits_{I=0}^m c_I \Big\langle \mbox{Tr}({\bf V} Q J_0 \star
{\bf V} Q J_0 \star) \Big\rangle_I\Big|_{\mbox{\scriptsize
\boldmath$V$}=\theta^4} = \mbox{One-loop result} - \frac{i}{2}\,
\frac{d}{d\log\Lambda}\qquad
\nonumber\\[3mm]
&& \times \sum\limits_f q_f^2 \sum\limits_{I=0}^m c_I\,
\mbox{Tr}\Big\langle\theta^4 \Big[y_\mu^*,
\Big[(y^\mu)^*,\log(\star) \Big] \Big]\Big\rangle_I -
\mbox{singular terms containing $\delta$-functions},\qquad\quad
\end{eqnarray}

\noindent
where it is necessary to subtract singularities of the expression in the second string of
this equation. Taking into account that trace of the $\theta$ commutators always vanishes, we
can replace $y_\mu^*$ with $x_\mu$ in this expression.

The term with $I=0$ ($c_0=-1$, $M_0=0$) corresponds to the case in
which the external lines are attached to the loop of ordinary
superfields $\Phi$ and $\widetilde\Phi$, while for $I\ge 1$ they
are attached to the Pauli-Villars  loop.
Therefore, it is possible to present Eq. (\ref{Double_Derivative})
in the form \cite{Shifman:2014cya}

\begin{eqnarray}
&& \mbox{One-loop result} + i\, \frac{d}{d\log\Lambda}
\sum\limits_f q_f^2\, \mbox{Tr}\Big\langle\theta^4 \Big[x_\mu,
\Big[x^\mu,\log(*) + \log(\widetilde *) \Big] \Big]\Big\rangle +
(PV)\qquad
\nonumber\\
&& - \mbox{singular terms containing $\delta$-functions},
\end{eqnarray}

\noindent where $(PV)$ denotes the contributions of
the Pauli-Villars loops.

According to Eq. (\ref{Double_Derivative}) the Adler $D$ function
(defined in terms of the bare coupling constant) is given by
integrals of double total derivatives exactly in the same way as the
$\beta$ function (defined in terms of the bare coupling constant)
in the Abelian case
\cite{Smilga:2004zr,Stepanyantz:2011bz,Stepanyantz:2011jy}. This
structure allows us to calculate one of the loop integrals
analytically and relate the result to the anomalous dimensions
of the matter superfields. To this end it is convenient to
rewrite Eq. (\ref{Double_Derivative}) in a different form.

Let us consider the sum of the digrams with $n$ vertices on the
matter loop to which external lines are attached. As in the
previous section this sum is denoted by the subscript $n$. Then,
after simple transformations (see Ref. \cite{Stepanyantz:2011jy}
for details), the right-hand side of Eq. (\ref{Double_Derivative})
(for simplicity, without the one-loop contribution) can be
equivalently presented in the form

\begin{eqnarray}\label{Single_Derivative}
&&\hspace*{-5mm} -\frac{d}{d\log\Lambda}\sum\limits_f q_f^2
\sum\limits_{I=0}^m c_I \mbox{Tr} \Big\langle \theta^4
\Big[y_\mu^*,\, \frac{i}{2} [(y^\mu)^*,I_0] \star -
\sum\limits_{a+b+2=n} \frac{(b+1)(\gamma^\mu)^{c{\dot d}}}{n}
(I_1)_c
(\star)_a (\bar I_1)_{\dot d} (\star)_b \Big]\Big\rangle_{I,n}\nonumber\\[3mm]
&&\hspace*{-5mm}  - \mbox{singular terms containing
$\delta$-functions},
\end{eqnarray}

\noindent
where $(\star)_a$ and $(\star)_b$ (with $a+b+2=n$) denote the $a$-th and $b$-th terms in the Taylor expansion
of $\star\,$, respectively. The traces of the commutators in Eqs. (\ref{Double_Derivative}) and (\ref{Single_Derivative})
evidently vanish. Therefore, the result for the Adler function is completely determined by the singular
contributions. In the next section we will calculate them starting from Eq. (\ref{Single_Derivative}).

\subsection{Summation of singularities}
\hspace*{\parindent}\label{Subsection_Singularities}

We have proved that the Adler function (defined in terms of the
bare coupling constant) is given by integrals of double total
derivatives in the momentum space. It is important to note that
these integrals do not vanish due to singularities, which
originate from the identity

\begin{equation}
\frac{\partial}{\partial Q^\mu} \frac{\partial}{\partial Q_\mu}
\frac{1}{Q^2} = -2 \frac{\partial}{\partial Q^\mu}
\frac{Q^\mu}{Q^4} = - 4\pi^2 \delta^4(Q)
\end{equation}

\noindent (which is written in the Euclidean space after the Wick
rotation). Indeed, for a nonsingular function $f(Q^2)$, with a
sufficiently rapid fall off at infinity,

\begin{eqnarray}
 \int \frac{d^4Q}{(2\pi)^4}\,\frac{Q^\mu}{Q^4} \frac{\partial
f}{\partial Q^\mu} &=& \frac{1}{8\pi^2} \int\limits_{0}^\infty
dQ^2\, \frac{d f(Q^2)}{dQ^2}
= - \frac{1}{8\pi^2} f(0)\nonumber\\[4mm]
&=& - \frac{1}{8\pi^2} \int
d^4Q\,\delta^4(Q) f = - \int
\frac{d^4Q}{(2\pi)^4}\,\frac{\partial}{\partial
Q^\mu}\frac{Q^\mu}{Q^4} \cdot f\,.
\end{eqnarray}

\noindent Beyond the one-loop approximation all diagrams in which
the external lines are attached to the closed loops of the
Pauli-Villars fields give a vanishing contribution, because there
are no such singularities. One loop   in this
approach should be considered separately.

Thus, we must consider only diagrams in which two external lines
are attached to a single loop of the (non-regulator)  superfields $\Phi$ and
$\widetilde\Phi$. This case corresponds to $I=0$. (Let us recall
that $c_0=-1$, $M_0=0$.) The $I=0$ part of Eq.
(\ref{Single_Derivative}) can be reduced to a simpler expression

\begin{eqnarray}
&&\hspace*{-3mm} \sum\limits_f q_f^2 \frac{d}{d\log\Lambda}
\mbox{Tr}\Big\langle \theta^4 \Big[y_\mu^*,\,
i(e^{2V}-1)\frac{\bar D^2 D^2 \partial_\mu}{8\partial^4} * -
\sum\limits_{a+b+2=n} \frac{2(b+1)(\gamma^\mu)^{c{\dot d}}}{n}
(e^{2V}-1)\frac{\bar D^2 D_c}{8\partial^2} (*)_a\nonumber\\[4mm]
&&\hspace*{-3mm} \times (e^{2V}-1) \frac{\bar D_{\dot d}
D^2}{8\partial^2} (*)_b + i(e^{-2V^t}-1)\frac{\bar D^2 D^2
\partial_\mu}{8\partial^4} \widetilde * - \sum\limits_{a+b+2=n}
\frac{2(b+1)(\gamma^\mu)^{c{\dot d}}}{n} (e^{-2V^t}-1)\frac{\bar
D^2 D_c}{8\partial^2}\nonumber\\[4mm]
&&\hspace*{-3mm} \times (\widetilde *)_a (e^{-2V^t}-1) \frac{\bar
D_{\dot d} D^2}{8\partial^2} (\widetilde *)_b \Big]\Big\rangle_n -
\mbox{singular terms containing $\delta$-functions}.
\end{eqnarray}

It is easy to see that the contributions corresponding to the
superefield $\Phi$ (which contain $e^{2V}$ and $*$) are equal to
the ones corresponding to the superfield $\widetilde\Phi$ (which
contain $e^{-2V^t}$ and $\widetilde *$). To prove this statement
we make the substitution $V \to - V^t$ in the functional integrals
in Eq. (\ref{Angular_Brackets}). Then

\begin{equation}
\Omega \to -\Omega^t;\qquad W_a \to -(W_a)^t; \qquad V_{\rm Adj} \to
V_{\rm Adj}\,,
\end{equation}

\noindent where (for arbitrary $X$ which belongs to the Lie
algebra of the gauge group) $V_{\rm Adj} X \equiv [V,X]$. This
implies that $S_{\mbox{\scriptsize gauge}}$, $S_\Lambda$,
$S_{\mbox{\scriptsize gf}}$, and $S_{\mbox{\scriptsize ghosts}}$
remain unchanged. As a consequence, we obtain the required
statement. Thus, the expression under consideration can be written
as

\begin{eqnarray}\label{Massless_Single_Derivative}
&&\hspace*{-3mm} \sum\limits_f q_f^2 \frac{d}{d\log\Lambda}
\mbox{Tr}\Big\langle \theta^4 \Big[y_\mu^*,\,
i(e^{2V}-1)\frac{\bar D^2 D^2 \partial_\mu}{4\partial^4} * -
\sum\limits_{a+b+2=n} \frac{4(b+1)(\gamma^\mu)^{c{\dot d}}}{n}
(e^{2V}-1)\frac{\bar D^2 D_c}{8\partial^2} (*)_a\nonumber\\[4mm]
&&\hspace*{-3mm} \times (e^{2V}-1) \frac{\bar D_{\dot d}
D^2}{8\partial^2} (*)_b  \Big]\Big\rangle_n - \mbox{singular terms
containing $\delta$-functions}.
\end{eqnarray}

\begin{figure}[h]
\begin{center}
\includegraphics[scale=0.34]{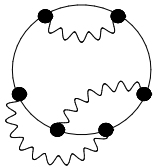}
\end{center}
\vspace*{-5mm} \caption{This diagram contains $p=2$ coinciding
momenta in the matter loop, which can lead to a singular
contribution. This is related to the fact that two cuts of the
matter line can make this diagram disconnected.}
\label{Figure_Example}
\end{figure}

Singular contributions appear if the derivative $\partial/\partial Q^\mu$ acts on
$Q^\mu/Q^4$. Such terms can come both from the first term of Eq.
(\ref{Massless_Single_Derivative}) and from the second term. In the
former case the derivative with respect to the momentum $Q^\mu$
corresponds to the commutator with $y_\mu^*$, and $Q^\mu/Q^4$ is
obtained from $\partial_\mu/\partial^4$. The latter possibility
requires the existence of the coinciding momenta in the matter loop.
Let us denote as
$p$ the number of the coinciding momenta in the matter loop.
An example of a diagram with $p=2$ is presented in Fig.
\ref{Figure_Example}.

Repeating the calculations of Ref. \cite{Stepanyantz:2011jy}, the
contribution of the first term in Eq.
(\ref{Massless_Single_Derivative}) can be rewritten as

\begin{eqnarray}\label{First_Contribution}
 - \sum\limits_f q_f^2\,
\frac{d}{d\log\Lambda}\mbox{Tr}\Big\langle \frac{\pi^2}{2}
\theta^4
* (e^{2V}-1)\bar D^2 D^2 \delta^4(\partial)\Big\rangle
= \frac{1}{2\pi^2} {\cal V}_4 \sum\limits_f q_f^2\cdot
\mbox{tr}\frac{d}{d\log\Lambda} G^{-1}\Big|_{Q=0}\,,
\end{eqnarray}

\noindent where $\mbox{tr}$ denotes the conventional matrix trace. By
definition, the function $G(\partial^2)$ is related to the
two-point Green function of the matter superfields as

\begin{equation}
\frac{\delta^2\Gamma}{\delta\Phi_j(x) \delta\Phi^{*i}(y)} =
\frac{\bar D_x^2
D_x^2}{16}G_i{}^j\delta^4(x-y)\delta^4(\theta_x-\theta_y)\,,
\end{equation}

\noindent where $\Gamma$ denotes the effective action of the
non-Abelian theory.

\begin{figure}[h]
\vspace*{2cm}
\hspace*{1.8cm}
\begin{picture}(0,0)
\put(1.5,-1.5){\includegraphics[scale=0.29]{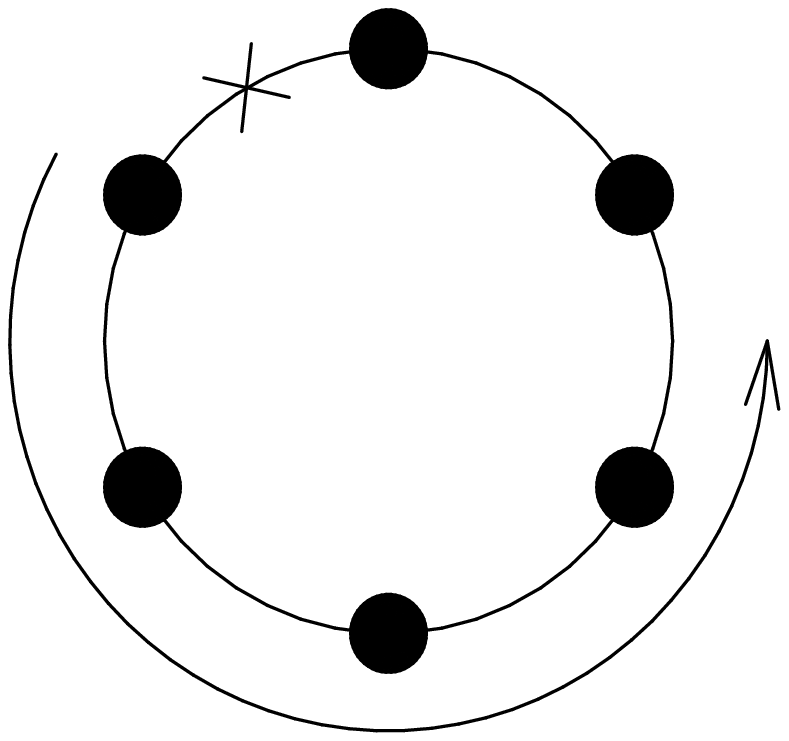}}
\put(4.7,-0.2){$+$}
\put(6.5,-1.5){\includegraphics[scale=0.29]{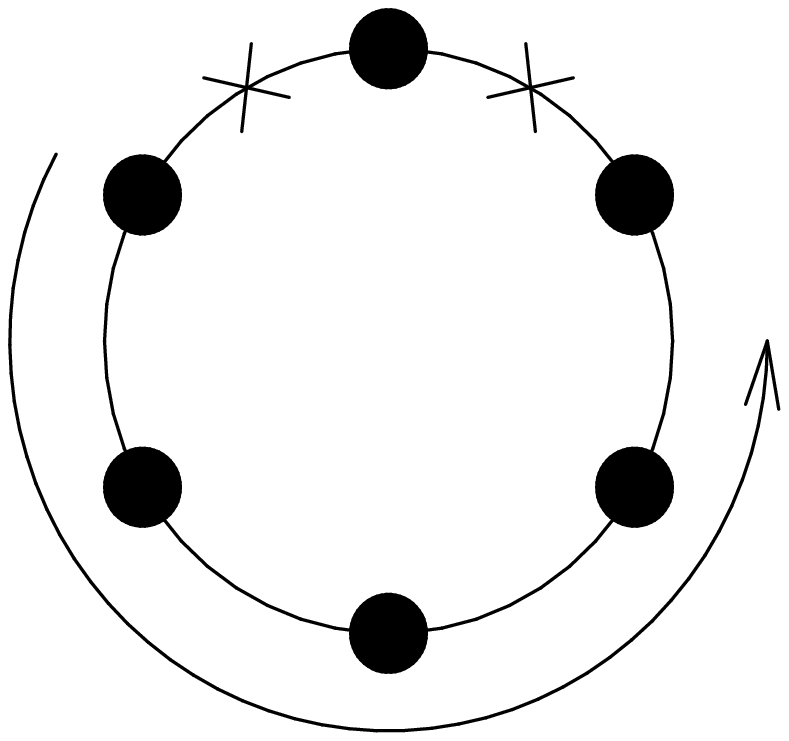}}
\put(1.5,1.1){${\displaystyle \frac{\bar D^2 D^2
\partial^\mu}{8\partial^4}}$}
\put(6.5,1.1){${\displaystyle \frac{\bar D^2 D_a}{8\partial^2}}$}
\put(8.2,1.1){${\displaystyle \frac{\bar D_{\dot b}
D^2}{8\partial^2}}$} \put(5.5,-0.2){$(\gamma^\mu)^{a{\dot b}}$}
\put(1.8,-1.2){$y_\mu^*$} \put(6.7,-1.2){$y_\mu^*$}
\put(0,-3){${\displaystyle G^{-1} = (1+\Delta G)^{-1} =
\sum\limits_{p=0}^\infty (-1)^p (\Delta G)^p }$}
\put(7.5,-3){${\displaystyle - \sum\limits_{p=1}^\infty
\frac{(-1)^p (p-1)}{p} (\Delta G)^p }$}
\put(2.7,-5){${\displaystyle 1+\sum\limits_{p=1}^\infty
\frac{(-1)^p}{p} (\Delta G)^p = 1- \log G}$}
\put(2.9,-1.5){\vector(0,-1){1}} \put(8,-1.5){\vector(1,-2){0.45}}
\put(2.9,-3.3){\vector(1,-1){1}} \put(8,-3.6){\vector(-1,-1){0.9}}
\put(10.0,-0.2){(for $p=6$)}
\end{picture}
\vspace*{5.3cm}
\caption{Obtaining the exact expression for the Adler function
by summation of singular contributions. The example presented
in this figure corresponds to the case $p=6$.}\label{Figure_NSVZ}
\end{figure}

The expression (\ref{First_Contribution}) is written formally,
because it gives integrals which are not well defined. The
well-defined integrals are obtained after adding contributions of
the remaining singularities. They appear from the second term in
Eq. (\ref{Massless_Single_Derivative}) if the matter loop has $p$
coinciding momenta ($p\ge 2$). They are situated between $p$ 1PI
subdiagrams (see an example in Fig. \ref{Figure_Example}
corresponding to $p=2$). Let us denote the numbers of vertices at
the matter line in these 1PI subdiagrams by $k_1, k_2, \ldots,
k_p$. The expression $Q^\mu/Q^4$ appears if there is the only 1PI
subdiagram between $D_c$ and $\bar D_{\dot d}$ in the second term
of Eq. (\ref{Massless_Single_Derivative}), see Fig.
\ref{Figure_NSVZ} in which the disks denote ($p=6$) 1PI
subdiagrams. This follows from the identities

\begin{equation}
\frac{\bar D^2 D_c}{8\partial^2} \cdot \frac{\bar D_{\dot d}
D^2}{8\partial^2} = i(\gamma^\mu)_{c{\dot d}}\,
\frac{\partial_\mu}{32\partial^4} \bar D^2 D^2\,; \qquad\quad
\frac{\bar D_{\dot d} D^2}{8\partial^2} \cdot \frac{\bar D^2
D_c}{8\partial^2}= - i(\gamma^\mu)_{c{\dot d}}\,
\frac{\partial_\mu}{32\partial^4} D^a \bar D^2 D_a\,.
\end{equation}

\noindent
Any other possibilities are excluded due to the equalities

\begin{equation}
\frac{\bar D^2 D^2}{16\partial^2}\cdot \frac{\bar D_{\dot b}
D^2}{8\partial^2}  = 0;\qquad\quad \frac{\bar D^2
D_a}{8\partial^2}\cdot \frac{\bar D^2 D^2}{16\partial^2} = 0.
\end{equation}

\noindent
Therefore, for $p\ge 2$ we obtain $p$ singular contributions inside the
second term in Eq. (\ref{Massless_Single_Derivative}), which correspond
to the following values of $a$ and $b$:
\begin{eqnarray}\label{AB}
&& a+1 = k_1; \qquad b+1 = k_2 + k_3 + \ldots + k_p;\nonumber\\[2mm]
&& a+1 = k_2; \qquad b+1 = k_1 + k_3 + \ldots + k_p;\nonumber\\[2mm]
&& \qquad\qquad\qquad\qquad\ldots \qquad\qquad\qquad\nonumber\\[2mm]
&& a+1 = k_p; \qquad b+1 = k_1 + k_2 +\ldots + k_{p-1}\,,
\end{eqnarray}

\noindent where $n = k_1 + k_2 +\ldots + k_p$. Then one can obtain
(see Ref. \cite{Stepanyantz:2011jy} for more details) that the
corresponding singular contribution differs from the contribution
of the first term in Eq. (\ref{Massless_Single_Derivative}) by the factor

\begin{equation}
-\frac{1}{p} \Bigg(\frac{k_1 + \ldots + k_{p-1}}{k_1 + k_2 + \ldots
k_p} + \frac{k_1 + \ldots + k_{p-2} + k_p}{k_1 + k_2 + \ldots k_p}
+ \ldots + \frac{k_2 + \ldots + k_p}{k_1 + k_2 + \ldots k_p}\Bigg) = -\frac{p-1}{p}
\end{equation}

\noindent
Therefore,  the coefficients in the sum of all singularities differ from the
coefficients corresponding to Eq. (\ref{First_Contribution}) by

\begin{equation}
1 - \frac{p-1}{p} =\frac{1}{p}\,.
\end{equation}

\noindent The whole contribution of diagrams with $p$ coinciding
momenta is proportional to $(\Delta G)^p$, where $\Delta G \equiv
G-1$. ($\Delta G$ corresponds to the sum of 1PI diagrams starting
from  one loop.) Then we compare the Taylor expansions

\begin{equation}
\log G = \log (1+\Delta G) = -\sum\limits_{p=1}^\infty
\frac{(-1)^p}{p} (\Delta G)^p\qquad \mbox{and}\qquad G^{-1} =
\frac{1}{1+\Delta G} = \sum\limits_{p=0}^\infty (-1)^p(\Delta
G)^p\,.
\end{equation}

\noindent
We see that after adding the singular contributions coming from the second term in Eq.
(\ref{Massless_Single_Derivative}) the result can be written as

\begin{eqnarray}\label{Whole_Contribution}
\frac{d\Delta\Gamma^{(2)}}{d\log\Lambda}\Big|_{\bm{V}=\theta^4}
&=& \mbox{One loop} - \frac{1}{2\pi^2} {\cal V}_4 \sum\limits_f
q_f^2\cdot \mbox{tr}\,\frac{d\log G}{d\log\Lambda}\Big|_{Q=0}
\nonumber\\[4mm]
& =& \frac{1}{2\pi^2} {\cal V}_4 \sum\limits_f q_f^2\cdot
\mbox{tr}\,\Big(1-\frac{d\log G}{d\log\Lambda}\Big|_{Q=0}\Big)\,.
\end{eqnarray}

\noindent Obtaining $\log G$ in this way is illustrated in Fig.
\ref{Figure_NSVZ}. Unlike Eq. (\ref{First_Contribution}), Eq.
(\ref{Whole_Contribution}) leads to the well-defined integrals.

The derivative of $\log G$ with respect to $\log\Lambda$ (which
should be calculated at a fixed value of the renormalized coupling
constant $\alpha_s$) in the vanishing momentum limit gives the
anomalous dimension defined in terms of the bare coupling
constant,

\begin{equation}
\frac{d\log G}{d\log\Lambda}\Big|_{Q=0} =
\frac{d}{d\log\Lambda}\Big(\log (ZG)-\log Z\Big)\Big|_{Q=0} =
-\frac{\log Z}{d\log\Lambda} = \gamma(\alpha_{0s})\cdot 1\,,
\end{equation}

\noindent where $1$ is written in order to stress that in the
non-Abelian case the anomalous dimension is $N_c \times N_c$
matrix, where $N_c$ is a number of colors. The matrix trace gives
the factor $N_c$, so that according to Eq.
(\ref{Adler_Function_Final}) we finally arrive at

\begin{equation}
D(\alpha_{0s}) = \frac{3}{2} \sum\limits_f q_f^2\cdot  N_c
\Big(1-\gamma(\alpha_{0s})\Big).
\end{equation}

\section{${\cal N}=1$ SQCD in the conformal window}
\hspace*{\parindent}\label{confwind}

In this section we will study some consequences ensuing from the
exact formulas (\ref{m5}) and (\ref{m6}) in the conformal window
\cite{Seiberg:1994pq,Intriligator:1995au} \beq \frac 32 N_c < N_f
< 3 N_c\,. \eeq

\begin{figure}[h]
\begin{picture}(0,6.8)
\put(2.1,6.4){$\frac{D(Q^2)}{\frac 32 N_c \sum_f q_f^2}$}
\put(3.0,4.9){$\frac{3N_c}{N_f}$} \put(3.3,2.4){$1$}
\put(12.8,0.3){$Q^2$}
\put(3.7,0.8){\includegraphics[scale=1.58]{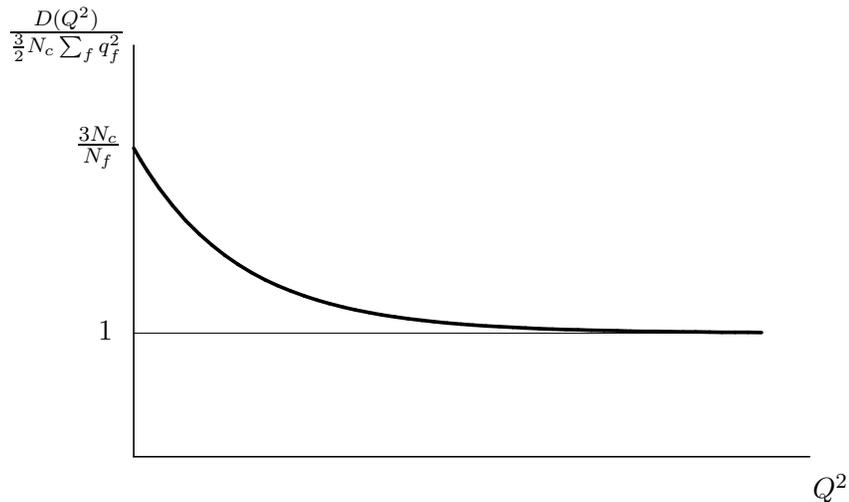}}
\end{picture}
\vspace*{-5mm} \caption{$D(Q^2)\cdot\left( \frac 32 N_c \sum_f
q_f^2\right)^{-1}$ versus $Q^2$.  The
horizontal lines corresponds to $N_f=3N_c$, i.e. the right edge of
the conformal window.
  }
\label{afcw}
\end{figure}

Inside this window SQCD flows to the conformal points: $\gamma=0$
in the ultraviolet (asymptotic freedom) and
\beq
\gamma_* =
-\frac{3N_c-N_f}{N_f}
 \label{82}
\eeq
in the infrared. The
equation (\ref{m5}) then implies that the Adler function \beq
D(Q^2) \longrightarrow \frac 32 N_c \sum_f q_f^2    \times
 \left\{
\begin{array}{c}
1\,,\qquad Q^2\to\infty\,,\\[3mm]
\frac{3N_c}{N_f}\,,\qquad Q^2\to 0\,.
\end{array}
\right.
\label{83}
\eeq
The $Q^2$ evolution of the Adler function in the conformal window is sketched in Fig \ref{afcw}.

The next interesting question is as follows: what new information
can be obtained by combining our exact formula (\ref{m5}) with the
Seiberg duality? We recall that the Seiberg duality connects with
each other  two distinct SQCD theories which flow to each other in
the infrared where their respective $\beta$ functions have fixed
points. If the number of colors in the original theory is $N_c$
the number of colors in its dual is $\widetilde N_c = N_f-N_c$.
The number of flavors in the original theory and its dual is the
same, and both have one and the same (global) flavor symmetry,
including the $SU(N_f)_L\times SU(N_f)_R$ factor. To introduce our
background $U(1)$ one can gauge a vector subgroup of the above
factor, for instance a diagonal subgroup of $SU(2)$. Then the
electric charges $q_f$ can be chosen as follows: \beq q(\Phi_1) =
q (\widetilde{\Phi}_2) =\frac{1}{\sqrt 2}\,,\qquad q(\widetilde
\Phi_1) = q ({\Phi}_2) =-\frac{1}{\sqrt 2}\,, \eeq with vanishing
charges for all other matter superfields. In this case \beq \sum_f
q_f^2 =1\,,\qquad \sum_f q_f = 0\,. \eeq

The dual theory has two couplings rather than one. In addition to
the dual gauge coupling it has a superpotential term \beq
{\mathcal W} = \lambda M_{f f'} \Phi_f\widetilde{\Phi}_{f'}\,,
\label{86} \eeq where $  M_{f f'}$ is the meson Seiberg field
\cite{Seiberg:1994pq,Intriligator:1995au}, and $\lambda$ is a
super-Yukawa constant. Both theories are superconformal in the
limits $Q^2\to \infty$ or 0. This means that the $\beta$ function
for $\lambda$ must have zero at  nonvanishing values of the coupling
constants.

Now, our exact formula gives $D(Q^2\to 0)$ both in the original and dual theories, and since
they are equivalent in the infrared, this should be one and the same value. Equation (\ref{83})
gives $D(Q^2\to 0)$ in the original theory. What about its dual?

Equation (\ref{m5}) is incomplete in the case of the dual theory.
Indeed, it takes into account only the $\Phi$ and $\widetilde\Phi$
matter fields. The $M$ field, being neutral with regards to the
gauge group $SU(\tilde N_c)$, is not neutral with regards to the
chosen flavor $SU(2)$ to which the background photon is coupled.
Thus, its loop must be included in an analog of Eq. (\ref{m5}).
Moreover, its superpotential interaction (\ref{86})  will
generates the anomalous dimension $\gamma_M$ of the the $M$
superfield. The anomalous dimensions of the $\Phi$ and
$\widetilde\Phi$  superfields acquire a matrix (in flavor)
structure. Their critical values are still given by (\ref{82})
with the replacement $N_c \to \widetilde N_c$.

Now, requiring that the infrared limit of the Adler functions in the original and dual theories
is the same, one should be able to
determine the critical value of $\gamma_M$.

A few words are in order here to explain why we think that our
analysis goes beyond perturbation theory. We are interested in the
effective Lagrangian for SQCD with photons expressed in terms of
various possible operators. The first operator is $F_{\mu\nu}^2$.
It appears perturbatively in one loop; non-perturbatively it could
appear as $f(\Phi^2/\Lambda^2)  F_{\mu\nu}^2$. However, any
function $f\neq 1$ would break an anomaly-free generalized  $R$
symmetry which is present in massless SQCD.

Nonperturbative effects can show up in the K\"ahler potential $K(
\Phi^+\Phi)$ in  the kinetic term for the matter fields. In the
effective Lagrangian it will be converted into a nonperturbative
term in $\gamma$. The relation between $\beta_\gamma$   and
$\gamma$ will remain the same as in (\ref{m5}), given the fact
that the conversion of $\Phi^+\Phi$ into $F_{\mu\nu}^2$ is the
exact one-loop Konishi anomaly.

Let us mention another general consequence from Eq. (\ref{m5}).
The Adler function $D(Q^2)$ is a physical quantity and, as such,
satisfies an appropriate dispersion relation. The equation
(\ref{m5}) implies that the anomalous dimension $\gamma(Q^2)$ must
satisfy the same dispersion relation.

\section{Conclusions}
\hspace*{\parindent}

Building on our previous publication \cite{Shifman:2014cya} we
expand our ideas on the exact Adler function in ${\mathcal N}=1$
SQCD. We present a very detailed proof of the master formula
(\ref{m5}) in perturbation theory. More exactly, we relate the
Adler function (defined in terms of the bare coupling constant)
with the anomalous dimension of the matter superfields (which is
also defined in terms of the bare coupling constant) to all orders
in the case of using the higher covariant derivative
regularization.

In essence, our relation is very similar to the NSVZ $\beta$
function and has a similar origin: all integrals for the $D$
function in the momentum space are, in fact, integrals of double
total derivatives. As a consequence, the singlet contribution,
which is proportional to $(\sum_f q_f)^2$, vanishes. The remaining
contribution does not vanish due to the existence of the integrand
singularities. We  proved that the sum of these singularities
gives the contribution proportional to the anomalous dimension.

It should be noted that the result obtained in this paper is
scheme-independent, because we consider the RG functions defined
in terms of the bare coupling constant. (These RG functions depend
on the regularization, but do not depend on the subtraction scheme
for a fixed regularization.) For the (scheme-dependent) RG
functions defined in terms of the renormalized coupling constant
the relation proposed in this paper is valid only in a special
subtraction scheme, which may be possibly constructed similarly to
the NSVZ scheme in ${\cal N}=1$ SQED by imposing certain boundary
conditions on the renormalization constants.

Our general arguments, both in \cite{Shifman:2014cya} and in this
paper (see also \cite{4}; the consideration in this paper is
complementary to ours), tell us that the relations between the
Adler function and the anomalous dimensions of the type (\ref{m5})
are valid beyond perturbation theory too. We discussed some
consequences of these relations in the conformal window. We
demonstrated that they allow one to determine the critical value
of the anomalous dimension $\gamma_M$ of the Seiberg $M$ field
which is present in the dual theory.

\section*{Acknowledgments}
\hspace*{\parindent}

We are very grateful to K. Chetyrkin, A. Kataev and D. Kutasov for valuable
discussions and communications. The work of MS is supported in part by DOE Grant
DE-SC0011842, and the work of KS is supported by the Russian
Foundation for Basic Research Grant 14-01-00695.

%\bigskip
%\bigskip

\newpage

\appendix

\noindent {\Large{\bf Appendix}}
\section{How to obtain subdiagrams and effective diagrams from the formal
expressions}
\label{Appendix_Subdiagrams}
\renewcommand{\theequation}{A.\arabic{equation}}
\setcounter{equation}{0}

Let us obtain the subdiagrams presented in Fig.
\ref{Figure_Subdiagrams} starting from the formal expressions
entering Eq. (\ref{Exact_Gamma}). First, we consider the
expression

\begin{eqnarray}\label{Subdiagrams1}
 \star \mbox{\boldmath$V$} Q J_0 &=& \frac{1}{1-I_0}
\mbox{\boldmath$V$} Q(I_0 + B_0 P) = \mbox{\boldmath$V$} Q B_0 P +
\frac{1}{1-I_0} I_0\mbox{\boldmath$V$} Q B_0 P
+ \frac{1}{1-I_0} \mbox{\boldmath$V$} Q I_0\qquad\nonumber\\[2mm]
& =& \mbox{\boldmath$V$} Q B_0 P + \star B\Big(
P\mbox{\boldmath$V$} Q B_0 P + \mbox{\boldmath$V$} Q P\Big),
\end{eqnarray}

\noindent where we use the notation

\begin{equation}\label{B0}
B_0 \equiv \left(
\begin{array}{cccc}
0 & 1 & 0 & 0\\[2mm]
1 & 0 & 0 & 0\\[2mm]
0 & 0 & 0 & 1\\[2mm]
0 & 0 & 1 & 0
\end{array}
\right)
\end{equation}

\noindent for the vertices with a single external
$\mbox{\boldmath$V$}$-line and no $V$ lines. Making simple
algebraic operations one can easily verify the identity

\begin{equation}\label{Identity_With_Pi}
P (\mbox{\boldmath$V$} B Q) P = P (\mbox{\boldmath$V$} B Q) (\Pi_+
P) + (P \Pi_-) (\mbox{\boldmath$V$} B Q) P,
\end{equation}

\noindent where the chiral projection operators $\Pi_\pm$ are
defined by

\begin{equation}\label{Chiral_Projectors}
\Pi_+ \equiv -\frac{\bar D^2 D^2}{16\partial^2};\qquad \Pi_-
\equiv -\frac{D^2 \bar D^2}{16\partial^2}.
\end{equation}

\noindent The equality (\ref{Identity_With_Pi}) has a simple
interpretation: one matter line coming from the vertex is chiral
and the second one is antichiral. In the first term the chiral end
of the right matter line is attached to the considered vertex, and
the chiral end of the left line is attached to the vertex in the
second term. Using Eq. (\ref{Identity_With_Pi}) we rewrite the
expression (\ref{Subdiagrams1}) in the form

\begin{equation}\label{Subdiagrams2}
\star \mbox{\boldmath$V$} Q J_0  = \mbox{\boldmath$V$} Q B_0 P +
\star B P\mbox{\boldmath$V$} Q B_0 P + B \mbox{\boldmath$V$} Q P +
(\star -1)\Big((B \mbox{\boldmath$V$} Q)\Pi_+ P + \Pi_- (B
\mbox{\boldmath$V$} Q) P\Big),
\end{equation}

\noindent where we take into account that

\begin{equation}\label{Star_Identity}
\star-1 = \Big(\sum\limits_{k=0}^\infty (I_0)^k B\Big) P = \star
BP.
\end{equation}

\noindent Because the propagators are chiral (or antichiral),

\begin{equation}
P = \Pi_{+} P + \Pi_{-} P.
\end{equation}

\noindent Therefore, again using Eq. (\ref{Star_Identity}) for
transforming the last term and taking into account that $Q$ and
$B$ commute we obtain

\begin{eqnarray}
\star \mbox{\boldmath$V$} Q J_0  = \mbox{\boldmath$V$} Q B_0 P +
\star \Big(B P\mbox{\boldmath$V$} Q B_0 P + (B \mbox{\boldmath$V$}
Q)\Pi_{+} P\Big) + B \mbox{\boldmath$V$} Q \Pi_{-} P + \star BP
\Pi_{-} (\bm{V} Q) B P.
\end{eqnarray}

\noindent Presenting $BP$ in the last term as $(I_0-1)+1$, the
considered vertex can be finally written in the following form:
\begin{eqnarray}\label{Subbdiagrams_Formal_Expression}
&& \star \mbox{\boldmath$V$} Q J_0  = \mbox{\boldmath$V$} Q B_0 P
+ \mbox{\boldmath$V$} Q B \Pi_{-} P  + \star BP \Pi_-
\mbox{\boldmath$V$} Q (I_0-1)
\nonumber\\[3mm]
&& + \star \Big(B P(\mbox{\boldmath$V$} Q) B_0 P +B
(\mbox{\boldmath$V$} Q)(\Pi_+ P) + B(P \Pi_-) (\mbox{\boldmath$V$}
Q)\Big).
\end{eqnarray}

\noindent Two first terms in the first string contain a single
vertex. In the third one the factor $I_0-1$ cancels $\star$ in
expressions like (\ref{Exact_Gamma}). These terms should be
considered separately. The terms in the second string give the
subdiagrams presented in Fig. \ref{Figure_Subdiagrams}. More
exactly, the subdiagrams in Fig. \ref{Figure_Subdiagrams}
correspond to terms in the brackets. Really, the first of these
terms contains the left vertex $B$ attached to the sequence of the
propagator $P$, the vertex with a single external line
$\mbox{\boldmath$V$} Q B_0$ and the propagator $P$. Thus, this
term encodes the subdiagrams in the first two columns of Fig.
\ref{Figure_Subdiagrams}. The second term (in the brackets in the
second string of Eq. (\ref{Subbdiagrams_Formal_Expression}))
consists of the left vertex $B Q$ to which the external line
$\bm{V}$ is attached and the propagators with the left chiral end
(due to the projection operator $\Pi_+$). The last (third) term
gives the left vertex $B$ and the propagators $P\Pi_-$, which have
the right chiral end to which the external line ($\bm{V} Q$) is
attached.

Let us apply Eq. (\ref{Subbdiagrams_Formal_Expression}) to the
calculation of various parts of Eq. (\ref{Exact_Gamma}). Let us
start with the diagrams in which the external lines are attached
to different matter loops. As we discuss in Section
\ref{Subsection_Two_Matter_Loops}, they include the expression
\begin{eqnarray}\label{Two_Matter_Loops}
&& \mbox{Tr}(\star \bm{V} Q J_0) =
\mbox{Tr}\Big\{\mbox{\boldmath$V$} Q B_0 P + \mbox{\boldmath$V$} Q
B \Pi_{-} P  - BP \Pi_- \mbox{\boldmath$V$} Q
\nonumber\\[3mm]
&& + \star \Big(B P(\mbox{\boldmath$V$} Q) B_0 P +B
(\mbox{\boldmath$V$} Q)(\Pi_+ P) + B(P \Pi_-) (\mbox{\boldmath$V$}
Q)\Big)\Big\},
\end{eqnarray}

\noindent where we use Eq. (\ref{Subbdiagrams_Formal_Expression}),
make a cyclic permutation in the third term and take into account
that $(I_0-1)\star = -1$. The terms in the second string of this
equation are discussed in Section
\ref{Subsection_Two_Matter_Loops}. Here we consider only the terms
in the first string of Eq. (\ref{Two_Matter_Loops}). Multiplying
the matrixes and calculating the trace one can easily see that
they give the vanishing contribution. Therefore, in Section
\ref{Subsection_Two_Matter_Loops} the terms in the first string of
Eq. (\ref{Subbdiagrams_Formal_Expression}) are not essential.

Let us proceed to the diagrams considered in Section
\ref{Subsection_One_Matter_Loop}. They are encoded in the
expression\footnote{For simplicity we omit the numerical factor
and the sum over $I$.}

\begin{equation}
\mbox{Tr}\Big\langle \star \bm{V} Q J_0 \star \bm{V} Q J_0
\Big\rangle,
\end{equation}

\noindent in which we substitute Eq.
(\ref{Subbdiagrams_Formal_Expression}). Taking into account that
$(I_0-1)\star =-1$ we obtain the following result:

\begin{equation}\label{ABC}
\mbox{Tr}\Big\langle \star \bm{V} Q J_0 \star \bm{V} Q J_0
\Big\rangle = A_2 + A_1 + A_0,
\end{equation}

\noindent where $A_2$ contains two operators $\star$, $A_1$
contains one operator $\star$, and $A_0$ does not contain $\star$
at all. It is easy to see that

\begin{eqnarray}\label{A}
A_2 &=& \mbox{Tr}\Big\langle \star \Big(B P(\mbox{\boldmath$V$}
Q) B_0 P +B (\mbox{\boldmath$V$} Q)(\Pi_+ P) + B(P \Pi_-)
(\mbox{\boldmath$V$} Q)\Big)\star \nonumber\\[3mm]
& \times& \Big(B P(\mbox{\boldmath$V$} Q) B_0 P +B
(\mbox{\boldmath$V$} Q)(\Pi_+ P) + B(P \Pi_-) (\mbox{\boldmath$V$}
Q)\Big)\Big\rangle
\end{eqnarray}

\noindent includes two copies of subdiagrams presented in Fig.
\ref{Figure_Subdiagrams}. This contribution corresponds to
diagrams $(a)$ -- $(e)$ in Fig. \ref{Figure_Effective_Diagrams}.
The contribution $A_1$ after some simple transformations can be
written as

\begin{eqnarray}\label{B}
 A_1 &=& 2\cdot \mbox{Tr} \Big\langle \star BP (VQB_0) P (VQB_0) P
+
\star (BVQ)(\Pi_+ P) (VQB_0) P \nonumber\\[2mm]
& +& \star B P (VQB_0) (P\Pi_{-}) (VQ) + \star (BVQ) (\Pi_{+} P
\Pi_{-}) (VQ)\Big\rangle.
\end{eqnarray}

\begin{figure}[h]
\vspace*{2.3cm}
\begin{picture}(0,0)
\put(0.5,0.06){\includegraphics[scale=0.35]{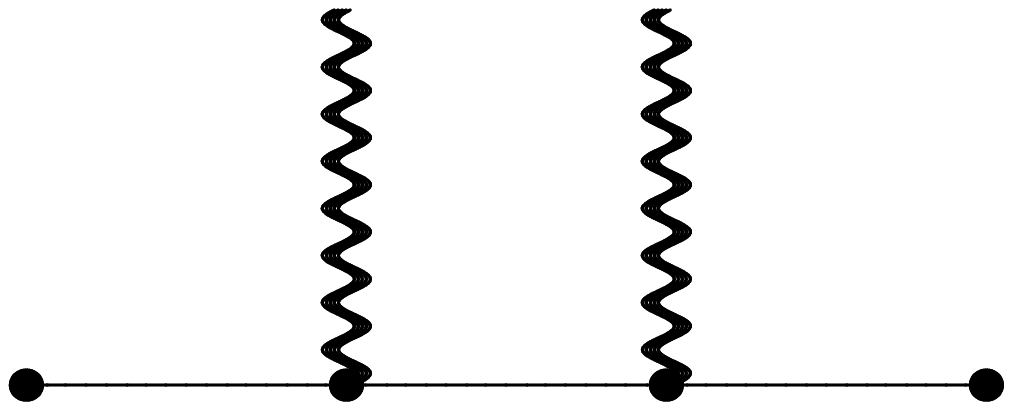}}
\put(5.4,0.06){\includegraphics[scale=0.35]{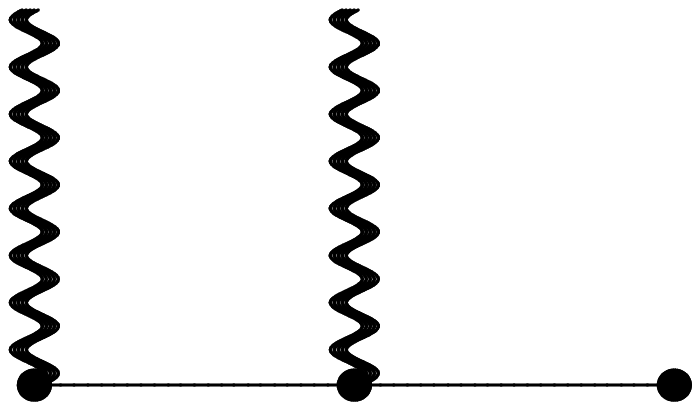}}
\put(8.9,0.06){\includegraphics[scale=0.35]{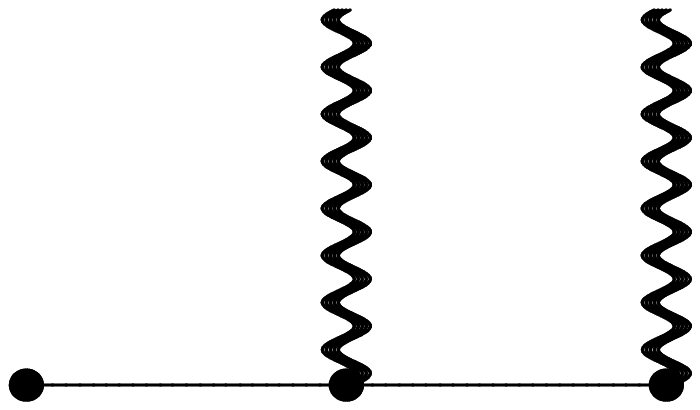}}
\put(12.5,0.06){\includegraphics[scale=0.35]{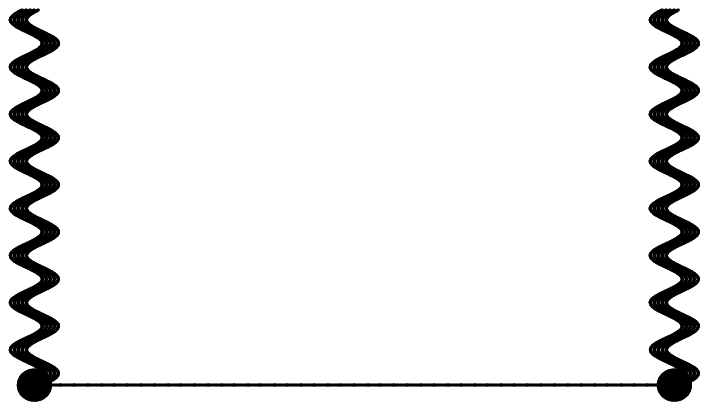}}
\end{picture}
\caption{Topology of subdiagrams which correspond to the
contribution $A_1$ in Eq. (\ref{ABC}) and form the effective
vertex in the diagram $(f)$ in Fig.
\ref{Figure_Effective_Diagrams}. In order to obtain all
subdiagrams it is necessary to take into account the projection
operators $\Pi_\pm$ in Eq. (\ref{B}).}
\label{Figure_Additional_Subdiagrams}
\end{figure}

\noindent (A lot of) subdiagrams which correspond to this term are
presented in Ref. \cite{Stepanyantz:2011jy} in Figs. 7-10
(together with subdiagrams proportional to $\bm{V}^2$ which comes
from the last term in Eq. (\ref{Exact_Gamma}).) These diagrams
have structure presented in Fig.
\ref{Figure_Additional_Subdiagrams}. However, it is also necessary
to point out chiral and antichiral ends of the propagators, which
should be made taking into account positions of the projection
operators $\Pi_{\pm}$ in Eq. (\ref{B}).

The contribution $A_1$ is graphically denoted by the diagram $(f)$
in Fig. \ref{Figure_Effective_Diagrams}. The last contribution
$A_0$ (that does not contains the operator $\star$) after some
transformations can be reduced to the one-loop expression

\begin{equation}\label{C}
A_0 = \mbox{Tr}\Big((\bm{V}QB_0) P (\bm{V}QB_0) P\Big).
\end{equation}

\noindent Certainly, the diagrams containing $\bm{V}^2$ vertex are
not present in this expression, because they come from the last
term in Eq. (\ref{Exact_Gamma}). Evidently, the angular brackets
in this case can be omitted due to the absence of internal gauge
lines.

\end{document}